\documentclass[a4paper,fleqn]{cas-dc}

\usepackage[numbers]{natbib}

\usepackage{tabularx}
\usepackage{listings} 
\usepackage{subcaption}

\def\tsc#1{\csdef{#1}{\textsc{\lowercase{#1}}\xspace}}
\tsc{WGM}
\tsc{QE}
\tsc{EP}
\tsc{PMS}
\tsc{BEC}
\tsc{DE}

\begin{document}
\let\WriteBookmarks\relax
\def\floatpagepagefraction{1}
\def\textpagefraction{.001}

\shorttitle{ }

\shortauthors{ }

\title [mode = title]{Large Language Model (LLM) for Software Security: Code Analysis, Malware Analysis, Reverse Engineering}                      
\tnotemark[1,2]


%

\author[1]{Hamed Jelodar}[                  
                        orcid=0000-0002-0713-3143]


\ead{h.jelodar@unb.ca}
 
\affiliation[1]{
    organization={Canadian Institute for Cybersecurity, Faculty of Computer Science, University of New Brunswick},
    addressline={\\46 Dineen Dr \#134}, 
    city={Fredericton},
    postcode={NB E3B 9W4}, 
    country={Canada}
}

\author[1]{Samita Bai}   
\ead{samita.bai@unb.ca`}

\author[1]{Parisa Hamedi}  
\ead{parisa.hamedi@unb.ca}

\author[1]{Hesamodin Mohammadian}  
\ead{h.mohammadian@unb.ca}

\author[1]{Roozbeh Razavi-Far}  
\ead{roozbeh.razavi-far@unb.ca}

\author[1]{Ali Ghorbani} 
\ead{ghorbani@unb.ca}

\cortext[cor1]{Corresponding author}



\makeatletter
\newcommand\thefontsize[1]{{#1\f@size pt\par}}
\newcommand{\showfont}{encoding: \f@encoding{},
  family: \f@family{},
  series: \f@series{},
  shape: \f@shape{},
  size: \f@size{}
}
\makeatother
\begin{abstract}
Large Language Models (LLMs) have recently emerged as powerful tools in cybersecurity, offering advanced capabilities in malware detection, generation, and real-time monitoring. Numerous studies have explored their application in cybersecurity, demonstrating their effectiveness in identifying novel malware variants, analyzing malicious code structures, and enhancing automated threat analysis. Several transformer-based architectures and LLM-driven models have been proposed to improve malware analysis, leveraging semantic and structural insights to recognize malicious intent more accurately.
This study presents a comprehensive review of LLM-based approaches in malware code analysis, summarizing recent advancements, trends, and methodologies. We examine notable scholarly works to map the research landscape, identify key challenges, and highlight emerging innovations in LLM-driven cybersecurity. Additionally, we emphasize the role of static analysis in malware detection, introduce notable datasets and specialized LLM models, and discuss essential datasets supporting automated malware research.
This study serves as a valuable resource for researchers and cybersecurity professionals, offering insights into LLM-powered malware detection and defence strategies while outlining future directions for strengthening cybersecurity resilience.
\end{abstract}


\begin{highlights}
\item A review of recent works on LLM applications in malware code analysis, summarizing current research trends.
\item First work exploring malware code analysis using LLMs, covering detection, generation, monitoring, reverse-engineering, and family analysis.
\item Investigation of how LLMs interpret source code semantics and structure to identify malicious behavior.

\item Insights into the role of LLMs in improving malware detection by recognizing and predicting code patterns.

\item Exploring the potential of LLMs to improve malware code analysis based on reverse-engineering process.

\end{highlights}


\maketitle

\section{Introduction}
In recent times, the growth of large language models (LLMs) has brought substantial changes to software security.
These smart language models are now being used to improve how we check code, find harmful software, and break down programs. With the extensive knowledge embedded within LLMs, security professionals can automate and refine the process of identifying vulnerabilities, understanding malicious behaviors, and deconstructing complex software systems. This integration not only accelerates analysis but also offers nuanced insights that were previously challenging to obtain through traditional methods. \\\\
Malicious code detection is a critical aspect of cybersecurity, as it helps protect computer systems and networks from various types of attacks. With malware’s increasing sophistication and complexity, traditional detection methods often struggle to keep pace. The advent of LLMs has opened up new possibilities for enhancing malicious code detection by leveraging the power of artificial intelligence (AI) and natural language processing (NLP) \cite{parla2024llm}. Several notable papers explore the application of transformers and LLMs across various fields, including malicious code detection and analysis of Portable Executable (PE) files\cite{jurecek2021representation,kunwar2024sok,tsfaty2022malicious},  
code classification ~\cite{zhao2025apppoet,fang2024llm}, code monitoring and inspection \cite{cani2014towards,nelson2024chatgpt,mohsin2024can,chen2024rmcbench}, and code generation \cite{ning2024mcgmark,cani2014towards,nelson2024chatgpt}.

For example, in \cite{xue2024poster}, the authors focused on evaluating LLMs for zero-shot learning in detecting malicious code, comparing code-specific models ( Code-llama \cite{roziere2023code}) with general-purpose models (Llama3\cite{llama3modelcard}, GPT-3.5 \cite{ye2023comprehensive}, and Claude\cite{Anthropic}). The study emphasized the effectiveness of prompts highlighting experience and maliciousness indicators over role-playing and noted that repeated questioning reduced LLM confidence. Moreover, in \cite{ning2024mcgmark}, the authors explored the risks of LLMs being exploited for malicious code generation, such as advanced phishing malware. They created MCGTEST, a dataset with 406 tasks, and developed MCGMARK, a robust watermarking method for tracing LLM-generated code. This approach, validated on the DeepSeek-Coder model ~\cite{guo2024deepseek}, achieved a success rate for embeddings while maintaining the code quality.\\

In \cite{qian2025lamd}, the authors focused on developing LAMD, a framework for Android malware detection and classification using LLMs. Their primary goal was to improve detection accuracy by analyzing malware behavior at multiple levels, ranging from low-level instructions to high-level semantics. This approach allowed for context-driven predictions and explanations, enhancing the understanding of the malware's operation. The researchers utilized JADX3 \cite{JTeam} to decompile the malware, concatenating all the pseudo source codes for input. This step was essential for ensuring a comprehensive analysis. In addition to GPT-4o-mini, they selected Gemini 1.5 Pro as another comparison model, praised for its ability to handle the longest context windows and its proven malware detection capabilities, making it a strong contender for effective malware classification. 

\begin{figure*}[htbp]
    \centering
    \includegraphics[width=0.96\textwidth]{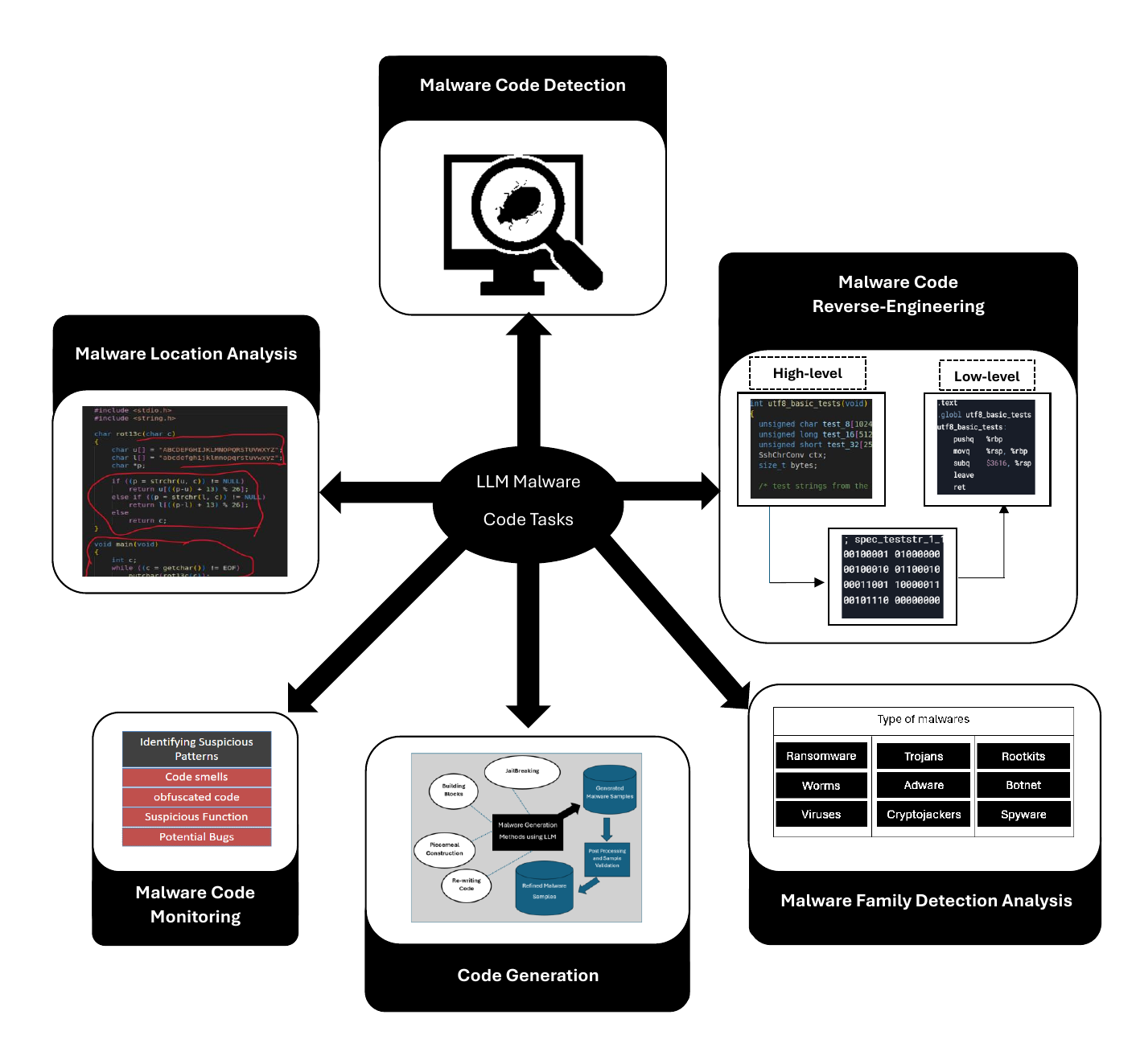}
    \caption{An overview of various malware code aspects explored in this research}
    \label{fig:research}
\end{figure*}

Recent research highlights the remarkable potential of LLMs in analyzing malware code, building on their proven success in understanding and interpreting such as. LLM models leverage advanced natural language processing techniques to interpret and analyze code patterns, enabling enhanced detection of vulnerabilities and malicious behaviors. LLMs can identify subtle nuances in malware coding techniques by utilizing pre-trained knowledge from vast datasets, facilitating improved threat detection and remediation strategies. As research in this area progresses, the integration of LLMs into cybersecurity frameworks promises to revolutionize how analysts approach malware analysis and response. In this paper, we aim to explore recent studies on how LLMs analyze the semantics and logic of source code to identify malicious behaviors. We highlight their capabilities in recognizing malware patterns and discerning logical flows indicative of malicious intent. We aim to provide an overview of advancements in LLM applications for malware detection and future research directions in cybersecurity. \\

\textbf{Research Motivation and Contributions:}Given the proven effectiveness of LLMs in various analytical domains, particularly natural language processing, and cybersecurity, we are motivated to explore their potential applications in malware code analysis. This study aims to examine how LLMs can be leveraged for detecting, analyzing, and understanding malware code through different models, datasets, and techniques. As illustrated in Figure~\ref{fig:research}, we investigate multiple aspects of malware code analysis using LLMs.
The primary objective of this work is to provide a comprehensive overview of LLM-based approaches in malware code analysis. In summary, this paper makes four key contributions:
\begin{itemize}
    \item A review of recent works on LLM applications in malware code analysis is presented, synthesizing current research findings and trends in the field.  
    
    \item To the best of our knowledge, this is the first work that investigates recent aspects of malware code analysis using LLMs, such as malware code detection, generation, monitoring, reverse-engineering, and family-code analysis.
    
    \item We explore how LLMs interpret the semantics and structure of source code to effectively identify malicious behaviors.  

 \item We investigate and present different future works and datasets that can be helpful for malware code analysis using LLMs.

\end{itemize}

\textbf{Evaluate and Analyze Relevant Research:} This study consider a comprehensive methodology for identifying pertinent research, encompassing the selection of data sources, search criteria, and the development of classification frameworks. The analysis is built upon an in-depth review of academic literature focused on malware detection in software code, with particular attention given to the most recent and widely cited papers that represent the forefront of research in the field. By critically examining these studies, we aim to uncover emerging trends, methodologies, and techniques that are shaping the evolution of malware detection, providing a snapshot of the latest advancements in cybersecurity. Furthermore, this study extends the discussion to the application of LLMs in the domain of malware code analysis, highlighting how these models can enhance detection, classification, and understanding of malicious code. Through this exploration, we aim to identify key opportunities for integrating LLMs into existing malware detection frameworks, offering insights into their potential to revolutionize malware analysis.\\

The rest of the paper is organized as follows. In Sections 2 and 3, we provide background on the core concepts of LLMs and related works. In Sections 4 to 9, we investigate and explore malware code analysis using LLMs, covering detection, generation, monitoring, reverse engineering, and family analysis. In Section 10, we discuss the most highlighted, practical, and recent LLM models with great potential for malware code analysis. Additionally, we examine extended datasets and LLMs for generating datasets for software code analysis tasks related to malware.

\section{Background}
\label{sec:background}
To improve the understanding and readability of concepts in this paper, this section provides background on key concepts related to methodologies, malware scenarios, and malware-code analysis tasks using LLMs.

\textbf{Reverse Engineering and Binary Analysis:} 
Reverse engineering and binary analysis are essential for dissecting and understanding malware ~\cite{omar2022reverse}. Since most malicious programs are distributed in compiled form without source code, security researchers rely on reverse engineering techniques to uncover their functionality. By analyzing binaries using tools such as IDA Pro \cite{eagle2011ida} and Ghidra ~\cite{eagle2020ghidra}, one can reconstruct the malware’s logic and identify malicious functions.

\textbf{Analyzing Malicious Software:} With the increasing volume and complexity of malware, analyzing malicious software has become a critical challenge. Malware analysis involves examining malicious code to understand its purpose, behavior, and impact on infected systems \cite{singh2020detection}. This process can be carried out using static or dynamic analysis methods, both of which play a crucial role in classifying malware families and enhancing cybersecurity defences.

\subsection{Core Concepts of LLMs}
LLMs are a big step forward in AI, improving language understanding and generation while enhancing capabilities in multiple fields \cite{yao2024survey}. LLMs operate on advanced deep learning frameworks, primarily utilizing transformer architectures and self-attention mechanisms. Their training relies on extensive datasets drawn from a wide range of textual sources, such as web content, published literature, and open-source codebases.\\

\begin{itemize}
    \item \textbf{Primally LLMs:} These LLMs are designed to handle a wide range of tasks rather than focusing on a single area. Currently, GPT and DeepSeek are popular examples of such models, excelling in logical reasoning, solving math problems, and generating creative writing. For example, in malware disassembly \cite{rong2024disassembling}, LLMs can be useful for analyzing malware by summarizing disassembled code and providing human-readable explanations of complex functions.

    \item \textbf{Prompt engineering:} Prompt engineering is a technique used to guide LLMs in generating accurate and relevant responses. It involves designing and refining prompts to help LLMs produce the desired outputs effectively \cite{sahoo2024systematic}. The structure and quality of prompts are critical in ensuring that the model generates accurate and relevant responses. Zero-shot and few-shot prompts are the most practical techniques. For example, a well-designed zero-shot prompt could ask an LLM to analyze obfuscated malware code and suggest deobfuscation techniques.

    \item \textbf{Training LLMs from Scratch:} Pre-training is a crucial step in developing LLMs, involving the exposure of models to vast amounts of unlabeled text data \cite{mckinzie2024mm1}. During this phase, the model learns grammar, context, and semantic relationships without specific task supervision. This foundational training enables the model to generalize across a wide range of applications before being fine-tuned for specialized tasks. However, pre-training LLMs on malware datasets, including malware repositories and threat intelligence feeds, significantly enhances their ability to assist in malware analysis \cite{rahali2021malbert}. 

    \item \textbf{Adapting Models for Specific Tasks:} Fine-tuning involves further training an LLM on domain-specific data to optimize its performance for particular applications. Unlike pre-training, which provides general language understanding, fine-tuning tailors the model to specialized tasks \cite{yao2024survey}. By training models on labeled datasets containing malware behaviors, code structures, and security reports, we can create specialized AI tools to detect, classify, and analyze threats \cite{setak2024fine}.

    Fine-tuning can be performed using supervised learning, reinforcement learning, or human feedback-based techniques. It allows organizations to customize AI models while reducing the computational burden of training from scratch. However, fine-tuning also introduces risks, such as overfitting or reinforcing biases in training data \cite{jin2020transferability}. 

    \item \textbf{Foundation Models and Architectures:}  Pre-trained models and transformer architectures form the backbone of modern LLMs. Transformers, introduced in the seminal paper “Attention Is All You Need,” revolutionized NLP by enabling efficient parallel processing and long-range contextual understanding \cite{vaswani2017attention}. These architectures use self-attention mechanisms to weigh different parts of input text, allowing for more coherent and context-aware responses.
\end{itemize}

\section{Related Work}
\label{sec:Relatedwork}
LLMs have numerous applications in malware analysis. Many studies have been conducted utilizing LLM-based approaches across various domains, such as cybersecurity, threat detection, reverse engineering, and digital forensics. Some research has specifically focused on surveying LLM applications in malware analysis \cite{khanllms, hossain2024malicious, zhang2024tactics}.

In \cite{xu2024large} the authors conducted a comprehensive review of the literature on LLM applications in cybersecurity, examining areas such as vulnerability detection, malware analysis, threat intelligence, and security policy automation. They analyzed 127 papers from leading security and software engineering venues, providing an in-depth overview of how LLMs have been utilized to address various cybersecurity challenges. The study primarily focused on research published between 2022 and 2023, categorizing the key characteristics, capabilities, and limitations of mainstream LLMs employed in the security domain.

In \cite{al2024exploring} the authors reviewed the use of LLMs for generating and detecting malware, explaining and organizing their findings clearly. They also provided strategies to reduce risks and defined key factors to measure how well these strategies worked. The metrics they created covered five areas: honeypots, text, and code-based threats, analyzing malware trends, and using existing malware to train LLMs. By creating example scenarios with data, they showed how effective these metrics were.\\

Based on our review of the existing literature, this study is the first to explore the application of LLMs to directly analyze and interpret malware code. This allows us to identify hidden patterns and take advantage of their capabilities for dynamic analysis and behavioral prediction.

\subsection{Malware Analysis: LLMs in Static \& Dynamic}
\begin{figure*}[h]
    \centering
    \includegraphics[width=17cm]{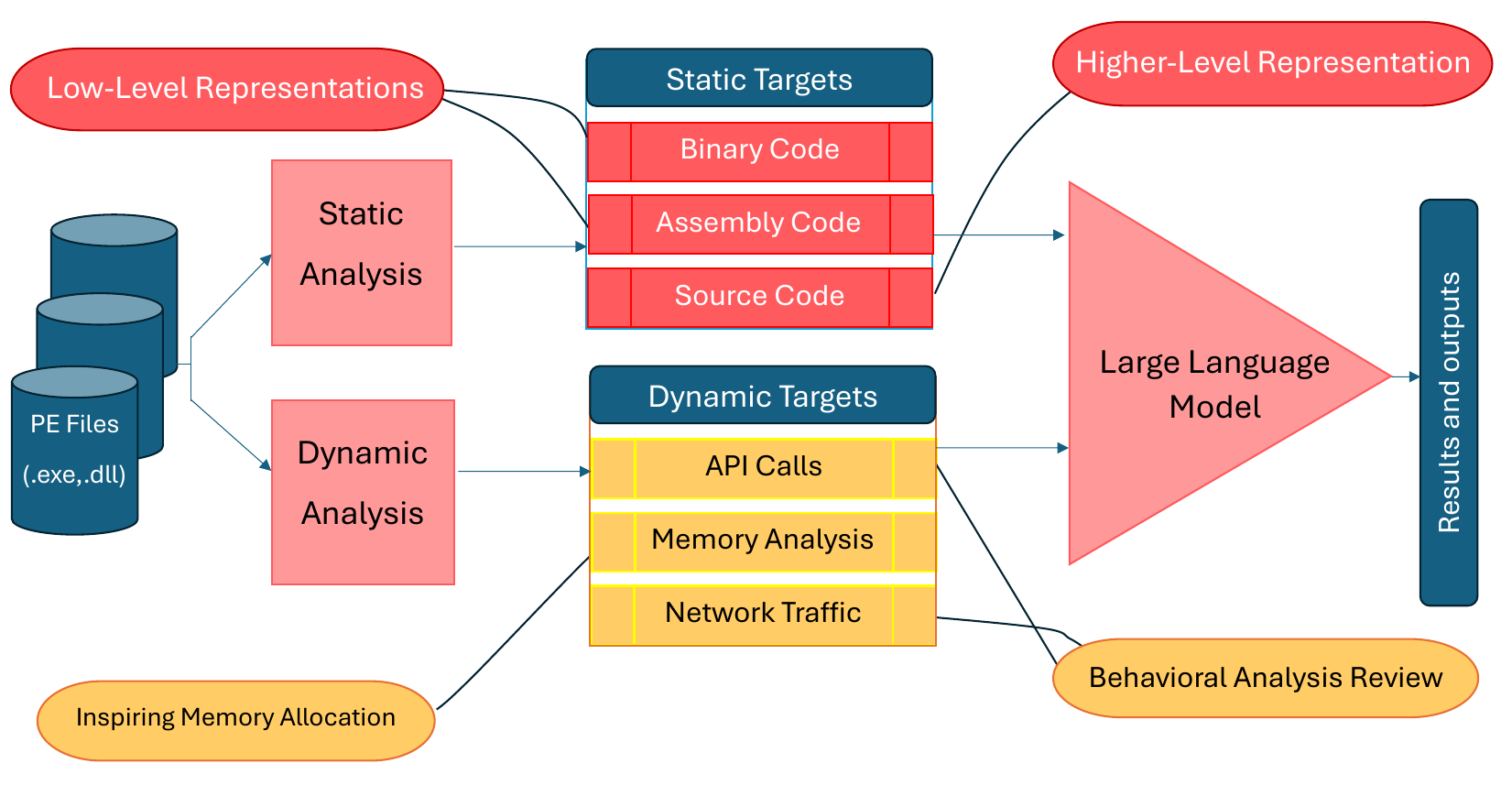}
    \caption{A general view of malware analysis based on static and dynamic ideas for feeding LLM.}
    \label{fig:mal_analysis}
\end{figure*}

This section examines the application of LLMs in malware code analysis, as illustrated in Figure~\ref{fig:mal_analysis}, emphasizing their role in secure coding, malicious code detection, and automated repair generation \cite{ lu2024malsight, islam2024enhancing, hoang2024novel, ccetin2024empirical}. Broadly, malware analysis using LLMs can be categorized into two main approaches: (1) Static Analysis and (2) Dynamic Analysis. These approaches provide a comprehensive framework for malware analysis powered by LLMs.

\subsection{Static Approaches}
Static malware analysis methods examine file hashes or binary code without execution to identify patterns that may disrupt computer performance \cite{lin2018efficient}.

\subsubsection{Static Approaches using LLM }
Different research efforts have explored the use of LLMs based on static malware analysis. For example, the authors in \cite{fujii2024feasibility}, focused on examines the application of LLMs to enhance static malware analysis, demonstrating high accuracy in describing malware functions. The authors employed a method involving static analysis workflows and utilized the OpenAI Codex model. They used a dataset comprising various malware samples to validate their findings \cite{lin2018efficient}. Also, other researchers, such as \cite{demirci2022static}, addressed the increasing threat of cyber attacks by proposing automated machine learning methods for detecting malicious code using advanced deep learning models. The authors employ stacked bidirectional long short-term memory (Stacked BiLSTM) and generative pre-trained transformer (GPT-2) models, utilizing assembly instructions from the .text sections of both malicious and benign PE files.

\subsubsection{Iterative Malware Summarization}
Some prior works have examined LLM models for malware summarization. For example, in \cite{lu2024malsight} the authors focused on addressing the challenges of generating accurate, complete, and user-friendly malware summaries by introducing MALSIGHT, a framework designed to improve malware behavior descriptions. They used and fine-tuned the CodeT5+ model \cite{wang2023codet5+} to enable transfer learning from source code summarization to decompiled pseudocode summarization. Their approach involved creating the first malware summary datasets, MalS and MalP, using LLM with manual refinement and training the specialized CodeT5+ model to iteratively summarize malware functions. Furthermore, the researchers in \cite{walton2025exploring} aimed to address the challenges of complex and time-consuming reverse engineering in Android malware analysis by introducing MalParse, a framework that leverages OpenAI’s GPT-4o-mini model for semantic analysis and categorization. Their approach enhances malware analysis by enabling automated, hierarchical code summarization at the function, class, and package levels, while also improving the efficiency of identifying malicious components.

\subsection{Dynamic Approaches}
Dynamic analysis differs from static analysis by focusing on identifying malicious patterns during execution rather than analyzing binary code. Unlike static methods, which are susceptible to evasion techniques, dynamic analysis does not require code disassembly \cite{or2019dynamic}. Its primary objective is to detect malicious activity in real-time while ensuring the security of the analysis platform. 

\subsubsection{Dynamic Approaches using LLM on API calls}
A handful of studies have investigated the dynamic approach and LLMs.  For example, in \cite{or2019dynamic}, the authors focused on improving malware detection by addressing the limitations of traditional dynamic analysis, which often struggled with obfuscated or unknown API calls. They used GPT-4 to generate explanatory text for each API call and embedded these explanations using BERT to create detailed representations of API sequences. These representations were then processed by a CNN-based detection model to identify malicious behavior.  Also, in \cite{or2019dynamic}  the authors introduced Nebula, a Transformer-based architecture for malware detection and classification, addressing the limitations of traditional dynamic analysis, which often relied on runtime APIs while neglecting heterogeneous data like network and file operations. Nebula processes diverse behavioral data from dynamic log reports using tokenization, filtering, normalization, and encoding components. 

\section{LLM for Malicious Code Detection}
\begin{figure}
    \centering
    \includegraphics[width=0.5\textwidth]{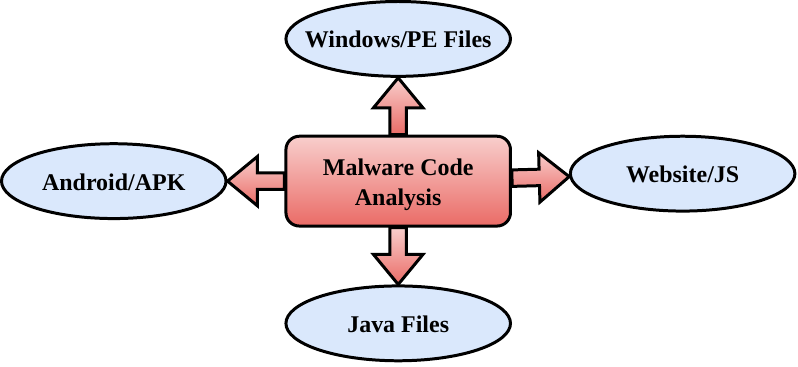}
    \caption{A general view of application of malicious code analysis based on different environments.}
    \label{fig:mal_code}
\end{figure}

\begin{figure*}[h]
    \centering
    \includegraphics[width=17.5cm]{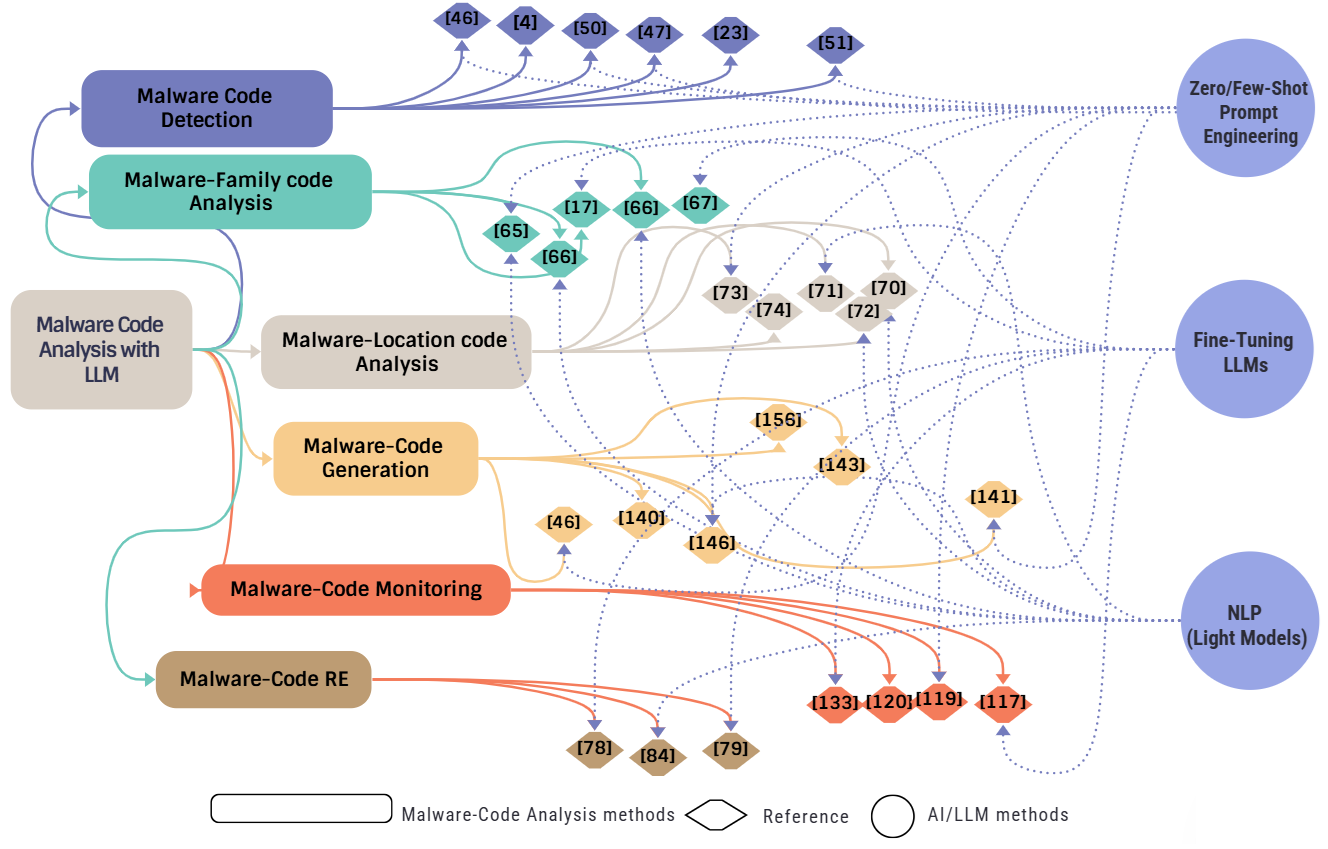}
    \caption{A bibliometric analysis and visualization of potential LLM and NLP methods for malware code analysis.}
    \label{fig:rel_models}
\end{figure*}

Recently, LLMs have provided a powerful solution for detecting both malicious and benign code by identifying patterns or malware family types. We can also see a relationship between malware code detection with different LLM methods in Figure \ref{fig:rel_models}.  \cite{al2024exploring}. 
Several studies have examined this method, which are summarized in Table~\ref{tab:my-tableRe}. Our analysis of previous research suggests that LLMs for malware code detection can be applied across four environments or data formats, as illustrated in Figure ~\ref{fig:mal_code}.


\subsection{Applications and Environments}
Previous studies indicate that the effectiveness of malware detection using static analysis and LLMs depends on the specific computing environment. Platforms such as Android, Windows, and Java each possess distinct characteristics that impact detection methodologies. For example, Android applications often involve complex interactions and permissions, necessitating tailored analysis techniques. Similarly, Windows executables and Java applications present unique structural and behavioral traits that require specialized approaches for accurate malware detection. Recognizing and understanding these environmental differences is crucial for developing effective platform-specific detection strategies.
\begin{table*}
\caption{Some impressive works in different applications of LLMs for malware analysis}
\label{tab:my-tableRe}
\begin{tabular}{lllll}
    \toprule
    \textbf{Ref}                                       & \textbf{Year} & \textbf{Application}                                                                               & \textbf{Models}                                                                                                                 & \textbf{Dataset}                                                                  \\ \midrule
    \cite{huang2024exploring}         & 2024      & Windows Malware Analysis                                                                           & GPT-3.5-Turbo                                                                                                                   & Created dataset with ChatGPT          \\ \midrule
    \cite{wang2024unmasking}          & 2024          & Windows Malware Analysis                                                                           & GPT4-Turbo                                                                                                                      & VirusTotal                                                                        \\ \midrule
    \cite{devadiga2023gleam}          & 2023          & Windows Malware Analysis                                                                           & GPT-3                                                                                                                           & 200000 PE files                                                                   \\ \midrule
    \cite{shestov2024finetuning}      & 2024          & Java Malware Analysis                                                                              & WizardCoder                                                                                                                     & CVEfixes                                                                          \\ \midrule
    \cite{hossain2024malicious}       & 2024          & Java Malware Analysis                                                                              & GPT-4                                                                                                                           & \begin{tabular}[c]{@{}l@{}}A Java Dataset using\\  Zero Trust Policy\end{tabular} \\ \midrule
    \cite{li2024enhancing}            & 2024          & Android Malware Analysis                                                                           & ChatGPT/GPT                                                                                                                     & Kronodroid dataset                                                                \\ \midrule
    \cite{walton2025exploring}        & 2024          & Android Malware Analysis                                                                           & GPT-4o-mini                                                                                                                     & 200 APK files                                                                     \\ \midrule
    \cite{wang2024liredroid}          & 2024          & Android Malware Analysis                                                                           & \begin{tabular}[c]{@{}l@{}}ChatGPT\\ ERNIE-Qwen\end{tabular}                                                                  &                                                                                   \\ \midrule
    \cite{zhao2025apppoet}            & 2024          & Android Malware Analysis                                                                           & \begin{tabular}[c]{@{}l@{}}GPT-4-1106-preview \\ Text-embedding-ada-002\end{tabular}                                       & AndroZoo                                                                          \\ \midrule
    \cite{pagano2025marvelhidedroid}  & 2025          & Android Malware Analysis                                                                           & ChatGPT/GPT                                                                                                                     &                                                                                   \\ \midrule
    \cite{wang2025llmdfa}             & 2025          & Android Malware Analysis                                                                           & \begin{tabular}[c]{@{}l@{}}gpt-3.5-turbo-0125\\ gpt-4-turbo-preview\\ gemini-1.0-pro\\ claude-3-opus\end{tabular} & TaintBench Suite                                                                  \\ \midrule
    \cite{roy2024phishlang}           & 2024          & Websites Malware Analysis                                                                          & MobileBERT                                                                                                                      & 25,796 Websites                                                                   \\ \midrule
    \cite{su2023bert}                 & 2024          & Websites Malware Analysis                                                                          & Bert                                                                                                                            &                                                                                   \\ \midrule
    \cite{roy2024chatbots}            & 2024          & \begin{tabular}[c]{@{}l@{}}Websites Malware Analysis\\ Generate malicious prompts\end{tabular} & Bert                                                                                                                            & \begin{tabular}[c]{@{}l@{}}1,255 phishing prompts\\ for websites\end{tabular}     \\ \midrule
    \cite{koide2024chatphishdetector} & 2024          & Websites Malware Analysis                                                                          & \begin{tabular}[c]{@{}l@{}}gpt-4-0314\\ gpt-3.5-turbo-0301\\ Gemini Pro\\ Llama-2-70b-chat-hf\end{tabular}                         & 1,000 malware/phishing sites                                                      \\ \bottomrule
    \end{tabular}
\end{table*}
\normalsize

\subsubsection{Static Malware Analysis for Android}
LLMs have emerged as powerful tools for malware detection and analysis, especially in Android applications \cite{pagano2025marvelhidedroid, li2024enhancing}. For example; in \cite{rahali2021malbert}, the authors stated that cyber threats and malware attacks have increased, posing significant risks to individuals and businesses. They emphasized the need for automated machine learning techniques to proactively defend against malware. To address this, they proposed MalBERT, a BERT-based Transformer model for static malware analysis. Their approach involved analyzing the source code of Android applications, extracting preprocessed features, and classifying malware into different categories. 

\subsubsection{Static Analysis of Malware in Java}
There is some impressive works that only focused on Java Malware files.  For example, in \cite{hossain2024malicious}, the authors utilized the Mixtral LLM model ~\cite{jiang2024mixtral} to identify malicious code in Java source files. They also leveraged NEO4J GenAI applications to analyze code structure, detect patterns, and uncover potential security threats. This approach enabled a more comprehensive understanding of malware behavior, improving detection accuracy and threat analysis.

\subsubsection{Static Malware Analysis of Websites}
Website malware refers to malicious software or harmful code that infects websites with the intention of damaging or exploiting the site or its visitors. This can include a variety of different attack types, including viruses, worms, Trojans, and scripts that steal data or inject unwanted content into a site. 
For example, in \cite{toth2024llms}, the authors focused on examining the security vulnerabilities introduced when LLMs, such as GPT-4, are used to generate PHP code for web development.
This analysis highlights the potential for exploitation, similar to how malware could exploit weak or improperly generated code in web environments. The research emphasizes the necessity of applying robust security measures when leveraging LLMs in web applications to prevent malicious activities.
 
\subsubsection{Static Analysis of Windows Malware Files}
There have been a few notable contributions to show applications of LLM for Malware analysis on PE files. For example in \cite{demirci2022static}, The authors focused on developing deep learning models, specifically Stacked BiLSTM and GPT-2, for detecting malicious code. They utilized GPT-2 models trained on assembly instructions extracted from both malicious and benign PE files. Their dataset included Win32 PE files from Windows operating systems, Commando VM v-2.0, and malicious x86 executables sourced from the sorel-20m database and the VirusShare website. In the next section, a detailed discussion is provided for malware detection strategies.

\subsection{Malware-Code Detection Strategies}
Malware detection has advanced with LLMs, particularly through fine-tuning pre-trained models on labeled malicious code. This method improves detection, especially for obfuscated or unseen code, by helping the model recognize complex malware patterns and domain-specific features.

\subsubsection{Fine-Tuning LLMs for Malware Code Detection}
Fine-tuning LLMs for malware code analysis involves tailoring a pre-trained model to specifically analyze malicious software code, helping to identify hidden threats within binaries, scripts, or other executable forms of malware \cite{lu2024malsight}. By training the LLM on a dataset that includes labeled samples of both benign and malicious code, the model learns to detect unusual patterns, code obfuscation techniques, and signature-less malware behaviors. Fine-tuning enables the model to better understand programming languages, malicious coding structures, and techniques such as polymorphism or encryption, which are often used to evade traditional detection methods. This allows for more effective identification of new and evolving malware strains.

Integrating fine-tuned LLMs with malware code analysis can significantly speed up the process of reverse engineering and threat detection \cite{yao2024survey} . Additionally, the LLM can offer context-driven insights or recommendations to help interpret obscure parts of the code, such as obfuscated functions or encrypted payloads. Over time, as more malware samples are analyzed, the fine-tuned LLM can evolve to recognize emerging techniques and improve the accuracy of predictions, contributing to a proactive defence against increasingly sophisticated malware attacks.

\subsubsection{Which models are highlighted for fine-tuning LLMs for malware code detection?}
Some works have been fine-tuned specifically for detecting malware using LLMs. The authors in \cite{lu2024malsight}, introduced MalT5, which was fine-tuned using the MalS dataset in combination with a benign pseudocode summarization dataset. Their underlying intuition was that the malicious semantic knowledge from malware source code summarization and the function patterns from benign pseudocode summarization, learned from these two datasets, could be transferred to generate malware pseudocode.

In \cite{wong2023attention}, the authors fine-tuned the BERT embedding and introduced a deep learning-based approach for behavior-based malware analysis. APILI identified API calls linked to malware techniques in dynamic execution traces. It used multiple attention mechanisms to connect API calls, resources, and techniques, integrating adversary tactics, techniques, and procedures through a neural network.

In \cite{demirci2022static}, the authors fine-tuned the BERT-base-uncased model to improve malware detection. They used Stacked BiLSTM and GPT-2 models to detect malicious code, creating language models from assembly instructions in benign and malicious PE files. Three datasets were formed: one for Document Level Analysis with Stacked BiLSTM, one for Sentence Level Analysis with various models, and a third for custom pre-training using unlabelled instructions.

In \cite{patsakis2024assessing}, they evaluated the power of modern LLMs for a well-defined malware analysis task: deobfuscating malicious PowerShell scripts. They used OpenAI's GPT-4 (gpt-4-1106) and Google’s Gemini "Pro" model, both accessed via their official APIs. Additionally, they considered two locally deployed LLMs: Meta’s Code Llama Instruct (with 34B parameters), based on Llama 2 and fine-tuned on 500 billion tokens of source code data, specifically designed for processing large chunks of codes.

In\cite{setak2024fine} The authors fine-tuned the CodeGen-Mono-350M model \cite{nijkamp2022conversational}, using Llama3-Instruct-8B as the teacher LLM. Llama3, being a highly capable code synthesizer, was ideal for bootstrapping the training dataset. 

\subsubsection{Pre-Training and Fine-Tuning}
Some prior works have examined the pre-training and fine-tuning steps for malware analysis based on LLM. In \cite{adamec2024development}, the authors introduced m-BERT, a deep learning model based on the BERT architecture, to address the gap in Python source code vulnerability detection. They curated a dataset by combining malicious code from open sources with benign code from PyPI. In the pre-training phase, a standard BERT model was trained with a masked language model objective to acquire generalized representations of Python code, resulting in m-BERT.  

\subsubsection{Few-Shot and Zero-Shot for malware code detection}
Few-Shot and Zero-Shot learning are advanced techniques used in malware analysis with LLMs, enabling them to identify and analyze malware with minimal task-specific training data \cite{zhang2024tactics}. Few-Shot learning allows models to recognize malware patterns from just a few examples, helping them generalize and detect new variants of malware in an ever-changing cybersecurity landscape. Zero-Shot learning takes this further by enabling malware detection without direct examples, relying on contextual understanding and prior knowledge \cite{chen2024rmcbench}.


\begin{figure}[h]
  
    \centering
    \includegraphics[width=0.45\textwidth]{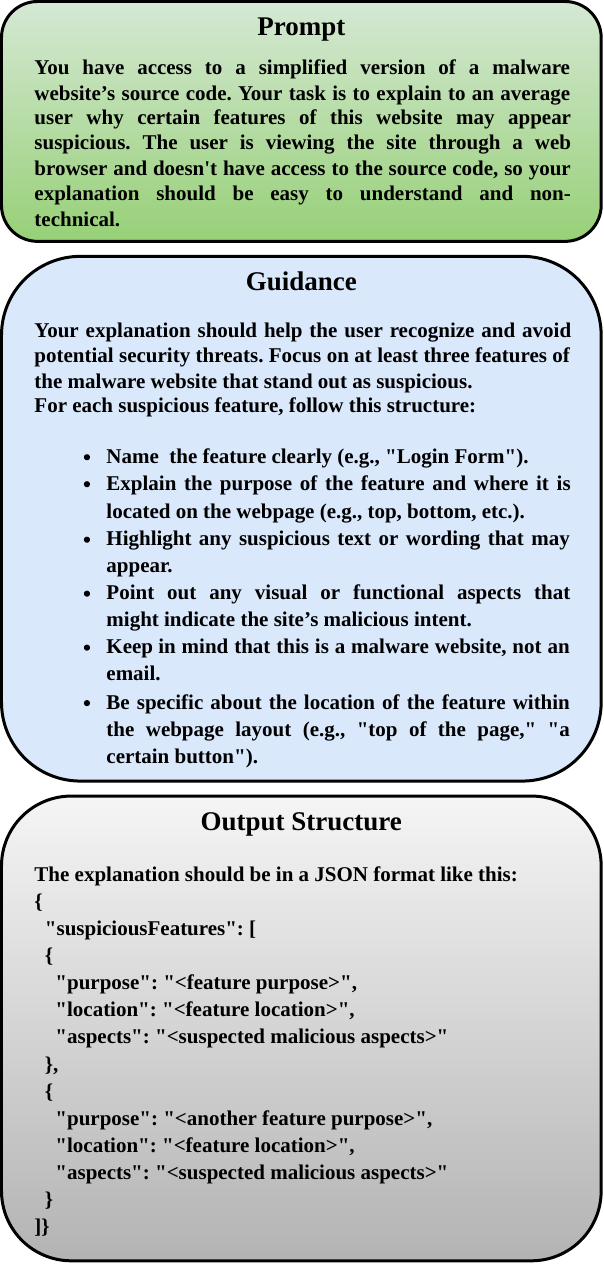}
      \caption{A sample prompt for malware code analysis using zero-shot approach for websites} 
          \label{fig:zero_website1}
\end{figure}

\begin{figure}[h]

    \centering
    \includegraphics[width=0.45\textwidth]{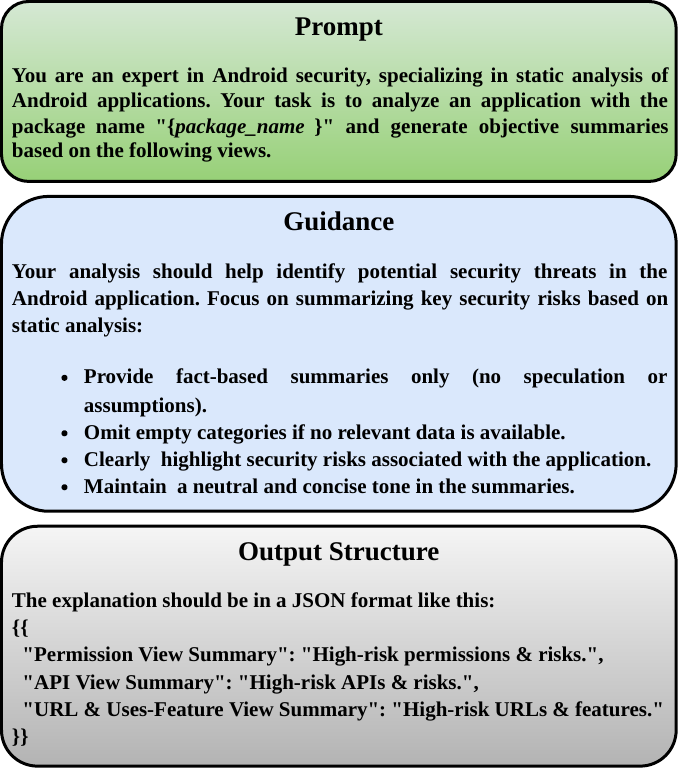}
        \caption{A sample prompt of malware code analysis using a zero-shot approach for Android apps} 
    \label{fig:zero_android}
\end{figure}

For example, in \cite{zhao2025apppoet}, different prompt engineering methods based on a GPT model to detect Android malware. They tested it on 11,189 benign and 12,128 malicious apps. The method first collected app features and created multiple views. In fact, it used prompt templates to guide the LLM in generating function descriptions and behavioral summaries. Figure~\ref{fig:zero_website1} and \ref{fig:zero_android} are sample prompts based on a zero-shot method for malware analysis on Android and websites. In more details, the prompt in the script instructs LLM model to perform a static security audit of an Android application, analyzing its permissions, APIs, URLs, and features for potential malware risks. It ensures the model generates structured, fact-based summaries highlighting high-risk elements, such as suspicious APIs or dangerous permissions, strictly formatted as JSON without speculation or unnecessary details.

In one such related work, in \cite{roy2024phishlang}, the authors developed a lightweight language model to detect phishing websites by analyzing their context based on prompt engineering techniques that called "Explainable Blocklisting". This system used GPT-3.5T to provide users with clear, context-driven explanations for why a website was flagged as phishing or malicious. 

\section{LLM for Malicious-Family Code Analysis}
Classifying malware into families is essential for understanding shared behaviors, identifying code reuse, and enhancing threat intelligence. This section presents an overview of techniques used for malware family classification, focusing on both non-LLM and LLM-based approaches. 

\subsection{Non-LLM }
In \cite{bao2024stories}, the authors proposed an NLP-based approach consisting of three key components: Information Collection, Semantic Embedding, and Familial Classification. More specifically, they employed a hierarchical attention mechanism to extract higher-order semantic information, which helps generate representation vectors for malware samples.

In a similar work \cite{sun2017malware}, authors conducted research on malware family classification using machine learning techniques to address the challenge of detecting polymorphic and variable malware variants, which traditional signature-based detection methods struggle to identify. They used a dataset of 65,536 malware samples from VirusShare, labeled based on the Kaspersky engine, and extracted features from three key aspects: bytecode features, assembler code features, and PE features. Through extensive feature selection and feature fusion processes, they refined the feature set and experimented with eight classifiers from Scikit-learn, including SVM, decision trees, and random forest.

In ~\cite{zhiwu2019android}, the authors enhanced CDGDroid to classify and characterize Android malware families using deep learning. They extracted control-flow (CFGs) and data-flow graphs (DFGs), encoded them into matrices for classification, and used n-gram sequences for characterization. 

\subsection{LLM-Based}
There are a few models for malware families based on transformers and LLM that we can mention, such as \cite{qian2025lamd,su2024codeart}. For example, 
In \cite{qian2025lamd}, the authors suggested a context-driven framework that used LLMs based on GPT-4omini/ChatGPT for Android malware detection, addressing challenges such as dataset biases, evolving threats, and the lack of explainability in traditional methods. They assisted manual audits by categorizing malware into six common families: Adware, Backdoor, PUA (Potentially Unwanted Applications), Riskware, Scareware, and Trojan, providing a structured approach to interpreting malware behaviors.

Also in \cite{su2024codeart}the authors propose a novel method for pre-training Transformer-based models when symbols are missing, using program analysis to extract contextual dependencies instead of relying on masked language modeling. Their model, a BERT-like Transformer with Regularized Multi-head Attention (RMA), incorporates both Masked Language Modeling (MLM) and Masked Dependence Modeling (MDM) for pre-training. This approach is applied to tasks like binary similarity, type inference, and malware family classification, using a dataset of 5484 samples from VirusTotal and VirusShare. The method aims to improve model performance in code analysis tasks without symbolic information.

\section{LLM for Malicious Code Location Analysis}
When malware is found in source code, pinpointing its location is crucial for effective removal and further analysis. Malware often hides itself within seemingly legitimate code, making it difficult to detect without careful examination. Malicious code typically resides in specific functions, modules, or files that exhibit unusual or suspicious behavior \cite{christodorescu2003static, samhi2022difuzer}. By analyzing the overall structure of the code, including function calls, imported libraries, and known malware patterns, we can identify these areas where the code performs potentially harmful actions.

\subsection{Non-LLM: NLP and Named Entity Recognition (NER) methods}
NLP techniques, such as Named Entity Recognition (NER), can be applied to detect entities such as function names, variable names, library imports, or system commands \cite{mohamed2024comprehensive}. Additionally, topic modeling can be used for malware analysis for opcode sequences, by treating a topic as a discrete distribution of opcodes \cite{medvet2016exploring,xu2023detime}. 

\subsection{LLM-based:  From Transformers to Large Models}
LLMs can be applied to identify the location of malware in source code by recognizing patterns and understanding the context of code. These models can detect suspicious behavior in code, such as abnormal API calls, network activity, or unusual file manipulations, which may indicate the presence of malware \cite{li2024llm}. By analyzing the semantic structure of the code, LLMs can highlight areas where the code performs potentially harmful actions, such as using system-level commands or interacting with external servers, which are typical indicators of malicious code. For example, by training LLMs on large datasets of known malware samples, these models can identify code patterns that indicate malicious intent \cite{huang2024spiderscan}.  These patterns could include unusual control flow, suspicious API calls, or anomalous interactions with system resources.

\section{LLM for Malicious Malware Reverse Engineering (RE)}
LLMs have emerged as powerful tools for malware detection by analyzing PE and other executable files. These models leverage static analysis and reverse engineering techniques to extract meaningful patterns from binary structures, API calls, and assembly instructions. By learning from vast datasets of benign and malicious samples, LLMs can effectively classify threats, detect obfuscated code, and enhance cybersecurity defences.

\subsection{ Non-LLM: Malware PE files and Reverse Engineering}
Reverse engineering of malware-infected PE files is crucial for understanding their behavior, attack vectors, and obfuscation techniques. This process involves deconstructing the binary file to examine its structure, API calls, and embedded payloads \cite{bhardwaj2021reverse}. Security researchers often use 1) disassemblers, 2) decompilers, and 3) debugging tools to analyze how malware interacts with system resources and executes malicious operations \cite{botacin2019revenge}, such as IDA PRO \cite{eagle2011ida} or Ghidra \cite{eagle2020ghidra}. Additionally, they can help deobfuscate code by predicting transformations used by malware authors, making it easier to recover the original malicious intent hidden within PE files.

\subsection{With LLM: Malware PE files and Reverse Engineering}
LLMs enhance reverse engineering by assisting in the automatic annotation of disassembled code, suggesting potential functions of unknown code segments, and identifying common patterns found in malicious binaries\cite{pordanesh2024exploring,tan2024llm4decompile}

From a reverse engineer's perspective, the next logical step is translating assembly language into a high-level language, which makes the program significantly easier to read, understand, and modify. Unfortunately, this task is highly challenging for any tool, as compiling high-level source code into machine code results in a significant loss of information. For instance, it’s impossible to determine the original high-level language of machine code just by examining it. While understanding specific characteristics of a compiler might assist a reverse engineer in identifying the machine code it produced, this approach is unreliable.\\

\subsubsection{LLM-Based: Low-level Code Generation}
The application of LLMs in low-level code generation, particularly for converting binary executables (such as PE files) into assembly, represents a significant advancement in reverse engineering, malware analysis, and security research \cite{mohseni2024can}. Traditional disassembly and decompilation techniques rely on static analysis tools, which often require extensive manual efforts to interpret obfuscated or packed binaries. However, LLMs can augment these processes by providing semantic understanding of machine instructions, enabling more efficient reconstruction of assembly code from raw binary data \cite{tan2024llm4decompile}. By utilizing pattern recognition and deep learning capabilities, LLMs can enhance the accuracy of disassembled output and even assist in reconstructing high-level control flow from low-level instructions. This has profound implications for malware analysis, as it can accelerate the identification of malicious logic, encoded payloads, and evasion techniques embedded within compiled binaries.

\subsubsection{LLM-Based: Analysis of a Packed Malware Binary}
A packed malware binary in the context of a PE file refers to a type of executable file that has been intentionally compressed or encrypted to hide its true contents. The purpose of packing is to make it harder for security software, such as antivirus programs or manual reverse engineers, to identify malicious code \cite{muralidharan2022file}. Suppose a cybersecurity team is investigating a PE file that is suspected of being a malware sample. Traditional disassemblers, such as IDA Pro or Ghidra, are able to partially decode the file, but much of the code remains obfuscated or packed, making it difficult to analyze. The team decides to apply a LLM to aid in converting the packed binary into assembly code, uncovering the malware’s behavior.

The analysis process begins by loading the packed binary into the investigation system. Due to the packing, which involves compressing or encrypting the file to obstruct analysis, manual unpacking would be time-intensive and require specialized expertise to identify the packing methods. To streamline this, an LLM is integrated with existing analysis tools. This model, fine-tuned on data involving packed executables and encryption techniques, can identify patterns associated with common malware packing methods, significantly accelerating the unpacking process.

Once integrated, the LLM utilizes its deep learning capabilities to decode the packed binary. It converts the raw binary into assembly code, which is far more readable and interpretable than the original machine-level instructions \cite{patsakis2024assessing}. This transformation is key to understanding the program’s behavior, as it provides a clearer view of the malware’s operations. Additionally, the LLM employs semantic analysis to detect obfuscation techniques such as encryption, anti-debugging, and code manipulation, which malware often employs to avoid detection. The decoded assembly code produced by the LLM offers a detailed and more accessible representation of the malware’s logic.

\lstset{
    language=C++,             
    basicstyle=\ttfamily,     
    keywordstyle=\color{blue},
    commentstyle=\color{green}, 
    stringstyle=\color{red},  
    numbers=left,             
    numberstyle=\tiny,        
    stepnumber=1,             
    frame=single,             
    tabsize=4,                
    showstringspaces=false,   
    breaklines=true           
}

 \begin{figure}[h]
    \centering
    \begin{minipage}{0.45\textwidth}
        \lstset{basicstyle=\ttfamily\footnotesize, frame=single}
        \begin{lstlisting}
int main() {
    system("gcc /tmp/malware_copy.c -o /tmp/malware_copy");
    system("/tmp/malware_copy &");
    system("openssl enc -aes-256-cbc -salt -in /tmp/important_file.txt -out /tmp/encrypted_file.txt");
    system("touch /tmp/.hidden_malware");

    return 0;
}
        \end{lstlisting}
        \captionof{figure}{Source Function (A)}
    \end{minipage}
    \hfill
    \begin{minipage}{0.45\textwidth}
        \lstset{basicstyle=\ttfamily\footnotesize, frame=single}
        \begin{lstlisting}
int __fastcall main(int argc, const char *argv, const char *envp) {
    system("gcc tmpmalwarecopy.c -o4 tmpmalwarecopy");
    system("tmpmalwarecopy");
    system("openssl enc -aes-256-cbc -salt -in tmpimportantfile.txt -out2 tmpencryptedfile.txt");
    system("touch tmp.hiddenmalware");
}
        \end{lstlisting}
        \captionof{figure}{Decompiled Function from a Malware PE File (B)}
    \end{minipage}
    \caption{Top-1 retrieved source function matched to a malware function query from a PE file.}
    \label{fig:similarfunction1}
\end{figure}

\subsection{Decompilers \& High-level Language Generation}
Decompilers are designed to transform binary code into high-level source code, often targeting specific languages and compilers. However, they typically do not produce fully compilable source code. The pseudo-code generated is often harder to interpret than human-written code, although it is generally easier to understand than raw assembly. These tools aim to translate binary code into high-level source code, tailored to specific languages and compilers \cite{chukkol2024vulcatch}.

\subsection{Deobfuscated and Obfuscated  on Malware codes (LLM and PE files)}
Hackers often obfuscate their scripts to avoid detection by security tools and analysts. This is commonly seen in malware, spyware, and other types of malicious code. When decompiling PE files, we may encounter obfuscated code as hackers use these techniques to hide their intentions \cite{choi2024chatdeob}. The goal is to make the code harder to analyze and reverse engineer, which complicates efforts to understand or neutralize the malicious payload. LLMs can help in automating the analysis of obfuscated code \cite{patsakis2024assessing}. These models can assist security professionals by quickly identifying patterns and functionality within obfuscated scripts, even when they are heavily disguised. LLMs can also generate insights and suggestions on how to reverse obfuscation, making it easier to detect and mitigate malicious behavior\cite{lin2024codecipher}. They can speed up the process of code analysis and even automate some tasks that would be time-consuming for analysts, improving efficiency in cybersecurity.

\begin{figure*}[htbp]
    \centering
    \includegraphics[width=0.8\textwidth]{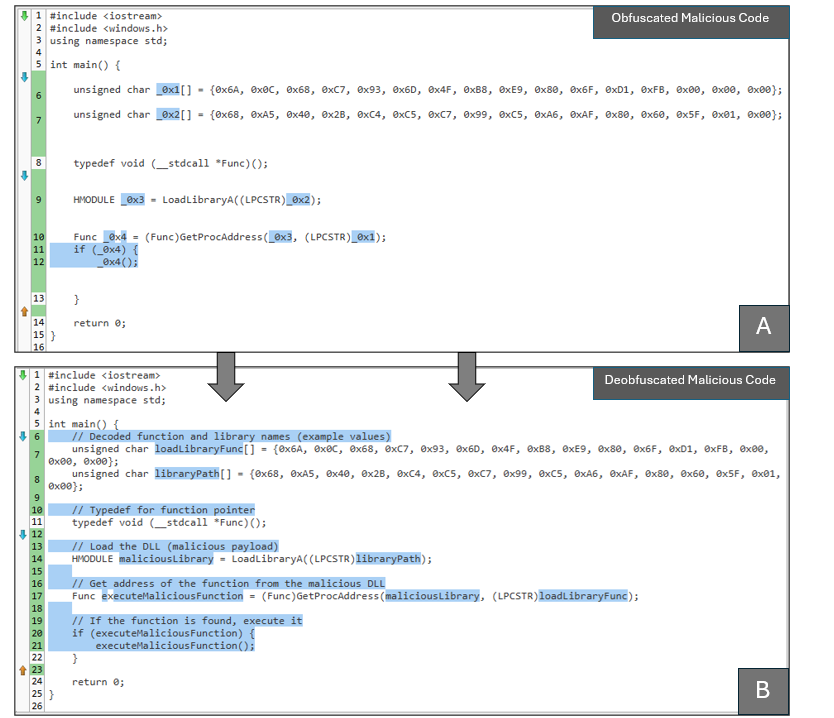}
    \caption{Example of the deobfuscation techniques used in malicious code.}
    \label{fig:deobfuscation}
\end{figure*}

Figure \ref{fig:deobfuscation}, illustrates two code examples that perform the same task — loading a malicious DLL and executing a function from it — but with different approaches. In \textit{Code B}, the variable names are clear, such as \textit{loadLibraryFunc} for the function to load and \textit{libraryPath} for the location of the DLL. This makes the code easy to read and analyze. In contrast, \textit{Code A} uses obfuscated names like \textit{\_0x1, \_0x2, and \_0x3}. This technique is designed to make the code more difficult to understand, helping it evade detection. The obfuscation in \textit{Code A} adds an extra layer of complexity, making it harder for security analysts or automated tools to quickly analyze the code. However, the core functionality remains unchanged, with both versions using the \textit{LoadLibraryA} and \textit{GetProcAddress}  functions. The key difference is that Code B sacrifices readability for stealth, which is a common tactic in malicious code to avoid detection. By leveraging LLMs, analysts can more efficiently tackle these obfuscated scripts and reduce the time it takes to identify and neutralize the threat.

\subsection{How well does the embedding model measure similarity between source functions and binary?}

The advancements in LLMs offer a unique opportunity to create a fully automatic and efficient system similar to the human disassembling process. Figure \ref{fig:similarfunction1} shows the top-1 most similar source function retrieved from a binary function query. This sample is considered positive, with a high degree of similarity.

In \cite{jiang2024binaryai}, the authors focused on developing BinaryAI, a binary-to-source SCA technique for intelligent binary source code matching. Its workflow included Feature Extraction, Embedding-based Retrieval, Locality-driven Matching, and Third-party Library Detection. The evaluation showed that the embedding model outperformed existing approaches in function retrieval

\subsection{Transformative Stages in Reverse Engineering and Pre-processing}
Regarding the Figure ~\ref{fig:mal_rev_eng}, the malware code and reverse engineering can be divided into three key stages: 1) data collection and reprocessing, where raw executable files are gathered and transformed into meaningful features; 2) pre-processing and reverse engineering; 3) tokenization and embedding. These stages are discussed below: 

\subsubsection{Data Collection and Preprocessing}
\begin{figure*}[h]
    \centering
    \includegraphics[width=1\textwidth]{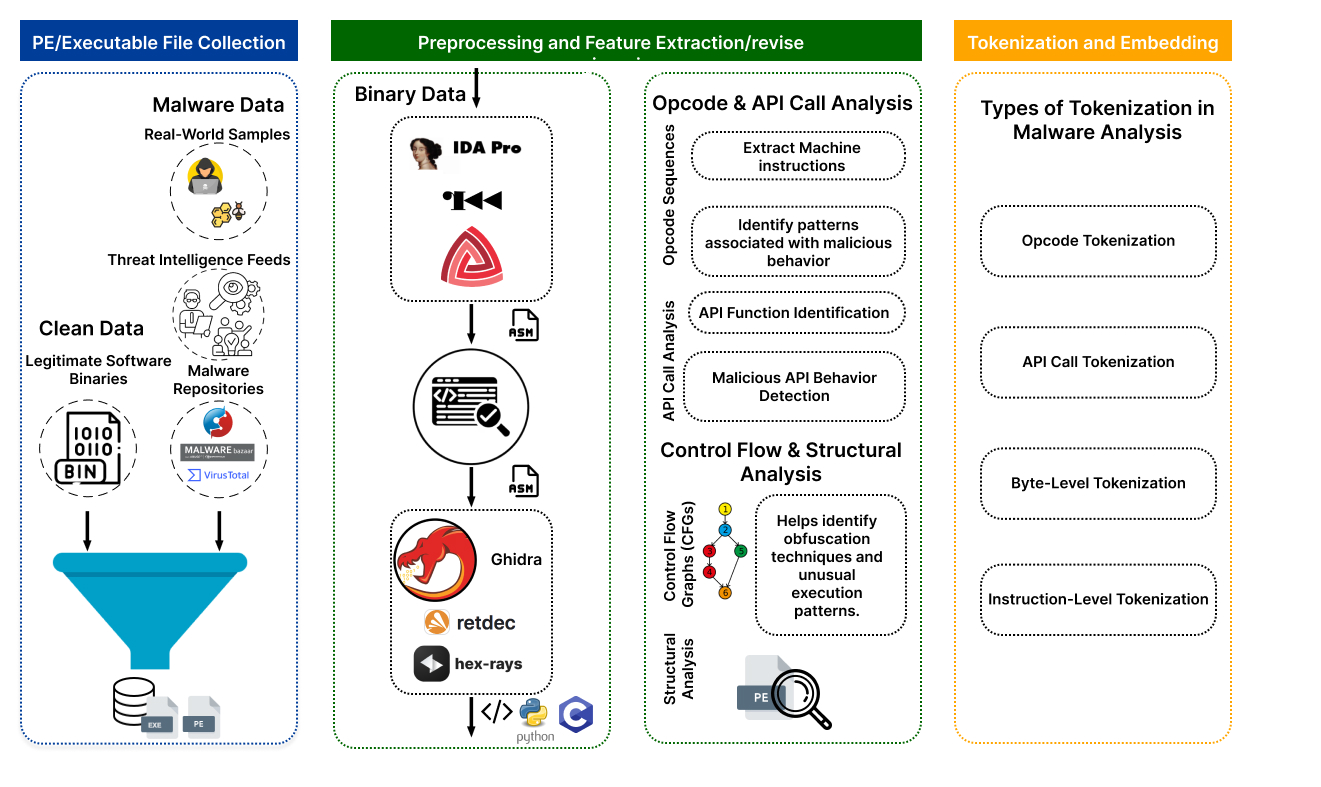}
    \caption{A general overview of pre-processing and reverse engineering for disseminating data on PE files.}
    \label{fig:mal_rev_eng}
\end{figure*}

For the first step in Figure~\ref{fig:mal_rev_eng}, we need to prepare and collect malware samples, specifically focusing on PE files. PE files are a standard binary format for executables, DLLs, and other system files on Windows operating systems. These files contain both machine code and metadata, making them a prime target for malware attacks. The prevalence of PE files in the Windows environment further enhances their attractiveness as a vector for malicious actors seeking to exploit vulnerabilities \cite{poudyal2018framework}.

\begin{figure}[h]
    \centering
    \includegraphics[width=0.48\textwidth]{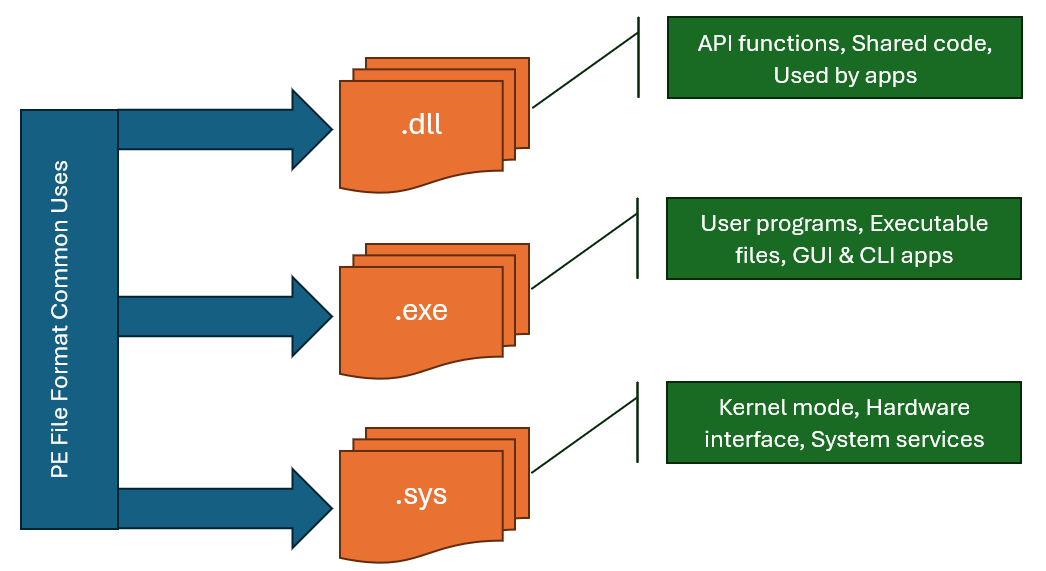}
    \caption{Types of PE files/formats.}
    \label{fig:pe_file}
\end{figure}

In particular, the structure of PE files often enables sophisticated techniques such as code injection, obfuscation, and packing, which can evade traditional detection methods. As shown in Figure~\ref{fig:pe_file}, there are various types of PE files, each serving different purposes, from executable files (.exe) to system driver files (.sys) and dynamic link libraries (.dll).  This high susceptibility to exploitation highlights the need for robust analysis techniques to detect malicious code within PE files.


To ensure a diverse dataset that includes both benign and malicious samples, PE files are gathered from multiple sources. Malware repositories, such as VirusTotal \cite{VirusTotal}, MalwareBazaar \cite{MalwareBazaar2024}, ANY.RUN \cite{anyrun}, and Hybrid Analysis provide labeled samples from public and private databases. Real-world samples are collected from infected systems or honeypots\cite{HoneyPot} designed to attract malware, including suspicious email attachments and drive-by downloads captured by web crawlers. Threat intelligence feeds \cite{ThreatIntelligenceFeeds1,ThreatIntelligenceFeeds2} contribute executable samples shared by security researchers and organizations in response to recent malware campaigns. Additionally, legitimate software binaries are sourced from official platforms to serve as a baseline for comparison. 

\subsubsection{Malware Reverse Engineering and Feature Extraction}
For the second step in Figure~\ref{fig:mal_rev_eng}, we need to utilize decompiling/disassembling tools to generate low-level or high-level code. Disassembly, the process of translating machine-readable bytes into human-readable assembly instructions, is crucial for binary reverse engineering tasks such as binary rewriting and vulnerability detection \cite{rong2024disassembling}. A major challenge lies in correctly identifying instruction boundaries, as the mixed storage of code and data or variable-length instructions can make it difficult to determine where instructions begin. This often leads to errors in identifying instruction boundaries and the instructions that follow. Tools like IDA Pro and Ghidra can heuristically convert binary code into C-like pseudo-code, but they often rely on heuristics and may struggle with complex or obfuscated binaries.

Opcode analysis and API call analysis are methods used to understand how software operates at different levels. Opcode analysis is low-level, focusing on the actual machine code and assembly instructions that the CPU executes, revealing details like memory manipulation and processor operations \cite{sun2019opcode, mowri2022application}. It’s crucial for reverse engineering and malware analysis. In contrast, API call analysis is high-level, tracking how a program interacts with the operating system and libraries through system calls, such as file operations or network communication \cite{wei2024mitigating}. While opcode analysis examines how a program runs at the processor level, API call analysis reveals how it interacts with its environment. Together, they offer a comprehensive view of a program’s behavior.

\subsubsection{Embedding and Tokenization}
In the third step of Figure~\ref{fig:mal_rev_eng}, it is essential to tokenize and embed the samples for feeding transformers or large language models (LLMs). Based on previous research \cite{gajdovsik2024machine}, there are different types of samples used for embedding, including opcode sequences \cite{kakisim2022sequential, khan2022op2vec}, API calls \cite{zhang2019feature, hu2025asdroid}, byte-level representations \cite{yousefi2018learning,abusnaina2023burning}, and instruction-level details\cite{korczynski2017capturing, mosli2021creating}. Each of these types provides a unique perspective on the data, and embedding them effectively helps the model understand and process the information efficiently for various tasks.

\subsection{Importance of LLMs for Malware PE File Analysis}

LLMs provide a groundbreaking solution to these challenges in malware code detection through static analysis of PE files. Unlike traditional models, LLMs possess contextual understanding, enabling them to analyze and interpret raw binary and code structures without extensive feature engineering. Their ability to generalize beyond predefined feature sets makes them highly effective against zero-day threats and sophisticated obfuscation techniques. Additionally, LLMs enhance explainability and adaptability, addressing the transparency concerns often associated with traditional machine learning models \cite{LLMforCyber}. Table \ref{tab:llm_nonllm} presents a comparative analysis of various non-LLM-based approaches in malware detection, specifically focusing on the static analysis of PE files. The studies listed highlight a range of traditional machine learning and deep learning methodologies, each with distinct advantages and limitations. While these approaches demonstrate high accuracy, scalability, and efficiency, they also face critical challenges that limit their effectiveness in handling complex and evolving malware threats. Regarding the Table, key observations from the table reveal that conventional methods such as transfer learning, ensemble learning, and feature-based classifiers exhibit strong detection capabilities. However, they also suffer from major drawbacks, including feature representation limitations, and computational overhead .

 \begin{table*}
    \centering
    \caption{Highlighting the advantages of LLMs compared to Non-LLM models.} 
    \resizebox{\textwidth}{!}{
    \begin{tabular}{c c c c}
        \toprule
        &   &  &  \\
        \textbf{Study} & \textbf{Non-LLM Approach}  & \textbf{Cons} & \textbf{Why LLM might} \\
         &   &  & \textbf{be a better choice?} \\
         &   &  &  \\
        \midrule
        &   &  &  \\

        \cite{OptimizedDeepLearning}  & CNN  & Feature Representation Limits &  Contextual Awareness\\
          & + & Computational Cost & Zero-Day Detection \\
          & Inception architecture & Vulnerability to Adversarial Attacks & Explainability \\
          &  & Lack of Context Awareness &  \\
          &  &  &  \\

        \cite{Ensemble-BasedTransferLearning} & TL + EML  & Computational Complexity & Contextual Understanding\\      
         &  & Feature Representation Limits & Adaptability\\
         &  &  & Explainability\\
        &  &  &  \\

         \cite{EffectiveRansomwareDetection}&PE Header Analysis&Limited Behavioral Insight&Contextual Understanding\\
         & + & Dependence on Known Patterns &Adaptability to Novel Threats \\
         & YARA Rules Integration &  & Explainability\\
         &  &  &  \\

        \cite{DetectionofmaliciousPE}
        &DNA Encoding of Features&Computational Complexity&Direct Analysis of Code and Behavior\\
        
        &+&Scalability Concerns& Scalability and Efficiency\\
        
        &Bioinformatics Analysis&Interpretability&Adaptability\\
        &&&Explainability\\
        &  &  &  \\

        \cite{RansomwareExtraction}&Static Analysis of PE Headers&Potential for Evasion&Contextual Analysis\\
        &+&Limited Behavioral Insight&Adaptability\\
        &Feature Extraction&&Comprehensive Detection\\
        &  &  &  \\

        \cite{MalwareDetectioninPE}
        & Feature Extraction  & Feature Selection Complexity & Advanced Pattern Recognition\\
        & + & Model Generalization &Adaptability to New Threats \\
        & RF/DT/& Computational Overhead & Reduced Feature Engineering\\
        & Gradient Boosting and AdaBoost/&  &  \\
        & Bayes Theorem&  &  \\
        & + &  &  \\
        & Benchmark Evaluation &  &  \\
         &  &  &  \\

        \cite{SwarmOptimization}& Dataset Development  & Static Analysis Limitations & Contextual Understanding\\
        & + & Optimization Overhead & Adaptability\\
        & Feature Selection  &  & Reduced Need for Feature Engineering\\
        & via Swarm Optimization &  &  \\
        & + &  &  \\
        & KNN/NC/RF/ &  &  \\
        & GNB/SVM/DT &  &  \\
        & &  &  \\

        \cite{ComparisonofFeatureExtraction}& Feature Extraction  & Evasion Techniques & Contextual Understanding\\
        & + & Limited Behavioral Insight & Adaptability\\
        & LR/SVM/KNN/Trees/ &  & Reduced Feature Engineering\\
        &  &  &  \\
        &  &  &  \\

        \cite{StaticAnalysis}& feature extraction  & Limited Behavioral Insight & Contextual Understanding\\
        & + & Evasion Techniques & Adaptability\\
        & CNN &  & Reduced Feature Engineering\\
        &  &  &  \\

         \cite{StaticDetectionofRansomwareUsingLSTM} & Feature Extraction  & Evasion Techniques&Comprehensive Data Understanding\\
        & + & Limited Behavioral Insight &Automated Feature Extraction \\
        & LSTM Network&  & Scalability\\
        &  &  &  \\

\bottomrule
        
    \end{tabular}
    }
    \label{tab:llm_nonllm}
\end{table*}

\section{LLM for Code Monitoring/Inspection and Malware Prevention}

While code reviews/inspections aim to prevent vulnerabilities from being introduced into software during development, malware analysis is focused on understanding and mitigating threats that exploit vulnerabilities. Malware code analysis is a reactive process that involves examining malicious code after it has been identified. The goal is to understand how the malware works, how it exploits vulnerabilities, and how to defend against it.

\subsection{Identification of Potential Defects}
The integration of LLMs into software development has opened new avenues for enhancing the efficiency and accuracy of critical tasks, including code review and inspection. Traditionally, these processes rely on manual efforts and static analysis tools, which, while effective to some extent, often face challenges such as high false positive rates, poor generalization, and coarse detection granularity. LLMs, with their advanced NLP capabilities, present a transformative opportunity to address these limitations. They offer not only the ability to identify potential defects in code but also the capacity to provide detailed, human-readable explanations and actionable suggestions for remediation. This dual capability positions LLMs as a promising tool for streamlining code review and inspection processes while improving their reliability and comprehensiveness~\cite{hou2024large,wang2024software}.

Recent studies have demonstrated that LLMs outperform traditional static analysis tools in identifying security defects and other vulnerabilities during code review. For instance, GPT-4 has shown remarkable potential in detecting defects, especially in smaller codebases or those written by less experienced developers. However, these studies also highlight significant challenges, including the generation of verbose responses and inconsistency in outputs. Prompt engineering techniques, such as few-shot and chain-of-thought prompting, have been explored to enhance LLM performance, but the variability in response quality remains a barrier to their widespread adoption~\cite{wang2024software,yu2024security}.

\subsection{Malware Prevention with Vulnerability Analysis }
Beyond vulnerability detection, LLMs have found utility in several related tasks that can fall under the category of code review in the software development lifecycle, such as defect detection, test generation, and bug analysis. These applications illustrate the versatility of LLMs but also underline the need for targeted evaluations to understand their specific strengths and limitations in code review. This section aims to provide a concise overview of how LLMs are being applied to code review and inspection, focusing on their capabilities, challenges, and practical implications for software development.

Among the above-mentioned tasks, vulnerability detection is the most crucial and has gained a lot of attention as a use case for LLMs. Vulnerability detection is a vital process in software development, focusing on identifying weaknesses in code that could be exploited by attackers. This task is closely tied to standards like Common Vulnerabilities and Exposures (CVE) and Common Weakness Enumeration (CWE), which provide a framework for categorizing and documenting security flaws. While traditional tools often struggle with precision and context-awareness,  LLMs  are emerging as a powerful alternative. By analyzing code with these established frameworks, LLMs can detect vulnerabilities more effectively and offer detailed explanations to help developers address security risks earlier and with greater accuracy.

One of the seminal works on using LLMs in vulnerability detection is done by Thapa et. al.~\cite{thapa2022transformer}. Their work explores the application of transformer-based models like BERT, GPT-2, and CodeBERT for identifying vulnerabilities in C and C++ source code. It introduces a systematic framework that integrates pre-training, fine-tuning, and inference to enable effective vulnerability detection. The framework leverages code gadgets—semantically related slices of code extracted based on data dependencies—to serve as inputs for the transformer models, enabling a structured approach to analyzing code vulnerabilities. The system flow involves pre-training on large text corpora, fine-tuning with labeled vulnerability datasets using added classification layers, and inference for predicting vulnerabilities in new code samples. This approach enables the models to adapt effectively to software-specific tasks while leveraging their pre-trained knowledge.

In one such related work by Szabo et al. ~\cite{szabo2023new}, authors explored using GPT models, particularly GPT-3.5 and GPT-4, to detect vulnerabilities in Angular-based web applications. It focuses on CWE-653 vulnerabilities related to the improper isolation of sensitive data handling.  The study introduced a sensitivity classification framework for data and tested the pipeline on large open-source Angular projects. 

Another work by Noever et al. ~\cite{noever2023can}, investigated the GPT-4's ability to detect and fix software vulnerabilities, comparing its performance with traditional tools like Snyk and Fortify. GPT-4 identified four times more vulnerabilities than Snyk across 129 code samples in eight programming languages and effectively reduced vulnerabilities by 90\% after applying its fixes. While demonstrating strong detection and self-correction capabilities, the study highlights limitations in handling large system-level codebases and occasional false negatives. The findings suggest GPT-4 can complement traditional static analyzers, offering a promising but still developing approach to software vulnerability management.

Another notable study that evaluated the performance of LLMs in vulnerability detection is presented in~\cite{gao2023far}. The authors introduced \textit{VulBench}, a comprehensive benchmark dataset that aggregates real-world and Capture-the-Flag (CTF) vulnerabilities. The study conducts a large-scale evaluation of 16 LLMs, including GPT-3.5, GPT-4, and several open-access models, comparing their effectiveness against deep learning-based methods and traditional static analysis tools. The evaluation considers both binary classification (whether a function is vulnerable) and multi-class classification (identifying specific vulnerability types). Results show that while LLMs outperform traditional deep learning models and static analyzers in controlled settings, their effectiveness declines in real-world scenarios requiring broader code context. The study also highlights the impact of few-shot prompting, showing that additional context improves detection rates. However, dataset quality remains a major challenge, as existing vulnerability datasets often contain inconsistencies. The findings suggest that while LLMs hold promise for vulnerability detection, they are best utilized in combination with static analysis and fuzzing techniques to enhance accuracy and reliability.

In~\cite{shestov2024finetuning}, the authors focused on fine-tuning LLMs for detecting vulnerabilities in Java source code. First, they start by selecting a suitable from open-source LLM models trained on large code corpora. They have selected WizardCoder~\cite{luo2023wizardcoder} which is compatible with transformer training frameworks and works better in answering questions with Yes or NO. Also, to address the challenge of performing classification tasks with LLM models they proposed a modified cross-entropy loss. The dataset they used in their work, was constructed from three open-source datasets: CVEfixes~\cite{bhandari2021cvefixes}, a Manually-Curated Dataset~\cite{ponta2019manually}, and VCMatch~\cite{wang2022vcmatch}. Their comprehensive evaluation study showed the fine-tuned WizardCoder model higher performance compared to previous CodeBERT-based models used for the vulnerability detection task.

Chan et. al.~\cite{chan2023transformer}, explored the use of transformer-based models for detecting vulnerabilities in code as developers write it, a process known as EditTime detection. Unlike traditional methods that analyze fully compiled code, this approach allows for real-time vulnerability detection, reducing the time between introducing and mitigating security risks. The authors trained models on a large dataset of vulnerable code snippets and experimented with zero-shot, few-shot, and fine-tuning techniques applied to CodeBERT, code-davinci-002, and text-davinci-003. Their study shows that fine-tuning provides a strong balance between accuracy and reliability, while zero-shot and few-shot learning offer more flexible but less precise detection. The research also highlights the risks of AI-generated code, showing that automated completions often introduce vulnerabilities. They also deployed their framework as a VSCode extension, significantly reducing vulnerabilities in developer-edited code and demonstrating the potential to integrate LLM-based security tools directly into the development workflow. Table~\ref{tab:vul_llm}, present a summary of works that incorporated LLMs in vulnerability detection tasks.

\begin{table*}
    \setlength{\tabcolsep}{5pt}
    \centering
    \caption{LLM-Based approaches for vulnerability detection.}
    \scalebox{0.8}{%
    \begin{tabular}{p{1cm}p{7cm}p{7cm}p{5cm}}
        \toprule
        \textbf{Ref} & \textbf{LLM Models} & \textbf{Language} & \textbf{Datasets} \\
        \midrule
        \cite{thapa2022transformer} & GPT-2, BERT, CodeBERT, DistilBERT, RoBERTa & C, C++ & VulDeePecker, SeVC \\
        \midrule
        \cite{szabo2023new} & GPT-3.5, GPT-4 & Angular & GitHub \\
        \midrule
        \cite{noever2023can} & GPT-4 & C, Javascript, Java, Go, C\#, PHP, Ruby & Codebase of Security Vulnerabilities \\
        \midrule
        \cite{gao2023far} & GPT-3.5, GPT-4, Llama-2, CodeLlama, Vicuna & C & VulBench \\
        \midrule
        \cite{shestov2024finetuning} & WizardCoder & Java & CVEfixes, VCMatch \\
        \midrule
        \cite{chan2023transformer} & CodeBERT, code-davinci, text-davinci & C++, Java, Javascript, C\#, Go, Ruby, Python & GitHub \\
        \bottomrule
    \end{tabular}%
    }
    \label{tab:vul_llm}
\end{table*}

\subsection{API Call Inspection}
Application Programming Interfaces (APIs) are essential to modern software development, enabling seamless interaction between diverse systems and services. As APIs grow in complexity, ensuring their correctness, reliability, and security has become increasingly challenging. Traditional API testing requires significant manual efforts to create and maintain comprehensive test suites, especially when dealing with complex data dependencies and evolving specifications. To address these challenges, recent research has explored the use of generative AI, particularly language models, to automate test generation, reduce human effort, and improve test coverage.

REST APIs, in particular, have become central to web services, microservice architectures, and the broader API economy. Automated REST API testing has emerged as a critical area of research, focusing on generating test cases from specifications like OpenAPI using strategies such as random input generation, evolutionary algorithms, and dependency analysis. However, existing tools often struggle with generating valid inputs, handling inter-request dependencies, and coping with incomplete or inconsistent specifications—leading to limited code coverage and missed faults. Recent work has explored solutions such as leveraging NLP to extract useful information from documentation, improving dependency inference, and incorporating dynamic feedback to create more robust and effective testing approaches ~\cite{kim2022automated}.

In ~\cite{kim2024leveraging}, the authors present RESTGPT, a novel approach that enhances REST API testing by extracting machine-readable rules from human-readable descriptions in OpenAPI specifications. Their framework, powered by GPT-3.5 Turbo, introduces a Rule Generator module that identifies four types of constraints: operational constraints, parameter constraints, parameter type and format, and parameter examples. Using carefully crafted prompts, RESTGPT improves API specifications by generating relevant parameter values and constraints, significantly outperforming existing techniques in both rule extraction accuracy and input value generation.

APITestGenie ~\cite{pereira2024apitestgenie} is an LLM-based framework designed to automatically generate executable REST API test scripts from natural language business requirements and OpenAPI specifications. By leveraging the contextual understanding capabilities of LLMs, the tool creates end-to-end integration tests simulating real-world scenarios, even across multiple API endpoints. APITestGenie achieved a test validity rate of up to 80\% through iterative prompt refinement and employed Retrieval-Augmented Generation (RAG) to handle large API specifications. While it enhances tester productivity, the tool is positioned as a supportive assistant requiring human validation before deployment in production environments.

In another work, LlamaRestTest~\cite{kim2025llamaresttest} introduces a resource-efficient approach to REST API testing by fine-tuning small language models, specifically the Llama3-8B model, to identify parameter dependencies and generate realistic test inputs. Unlike static specification-based methods, LlamaRestTest actively incorporates feedback from server responses during test execution to iteratively refine requests, ensuring higher accuracy and compliance with complex parameter constraints. Through fine-tuning and quantization, LlamaRestTest outperformed larger models and existing testing tools in both code coverage and fault detection, offering a scalable and cost-effective solution for dynamic and comprehensive API testing.


\subsection{Code Smells and Pattern Analysis}
Code smells indicate poor coding practices that increase software complexity and maintenance challenges, often leading to security vulnerabilities. While not immediate defects, issues like long methods, excessive coupling, and global variables can obscure logic and make exploitation easier. Studies show that smells such as long parameter lists and complex class structures correlate with security flaws ~\cite{elkhail2019relating}. In languages like JavaScript, security-related smells such as hardcoded credentials, excessive global variables, and improper error handling can lead to injection attacks, unauthorized access, and data breaches ~\cite{kambhampati2024characterizing}. Detecting and refactoring these smells early helps mitigate security risks and improve software robustness.

To address the limitations of traditional code smell detection, researchers have integrated machine learning and deep learning to enhance detection accuracy and reduce the subjectivity of heuristic-based methods. Unlike manual threshold-setting approaches, ML and DL models learn to differentiate between smelly and non-smelly code, mitigating performance instability. The use of LLMs in code smell detection has gained increasing attention in recent years, leveraging their advanced NLP capabilities and ability to understand complex software patterns. Traditional code smell detection methods relied on heuristic-based approaches, machine learning (ML), and deep learning (DL) models, which often struggled with subjectivity, high computational requirements, and limited generalization ~\cite{zhang2024comprehensive}. Recent studies have explored prompt-based LLMs, fine-tuned LLMs, and hybrid approaches that integrate expert toolsets to improve detection accuracy and efficiency.

A key development in this space is the exploration of Parameter-Efficient Fine-Tuning (PEFT) methods to adapt LLMs for method-level code smell detection ~\cite{zhang2024comprehensive}. Unlike full fine-tuning, PEFT techniques such as LoRA, prefix tuning, and prompt tuning have demonstrated comparable or superior performance while consuming fewer computational resources. However, findings suggest that small LMs (under 1B parameters) can sometimes outperform larger LLMs for specific method-level smell detection tasks. This suggests that task-specific adaptations, rather than model size alone, play a crucial role in optimizing LLM-based detectors.

Another promising approach integrates LLMs with expert toolsets for code smell detection and refactoring. The iSMELL framework employs a Mixture of Experts (MoE) architecture, selecting the most suitable expert tool for each smell type and using LLMs to refine refactoring suggestions ~\cite{wu2024ismell}. This hybrid approach addresses LLMs’ limitations, such as difficulty in repository-level knowledge retrieval, token constraints, and dynamic source code analysis. Experimental results show that iSMELL outperforms baseline LLMs and traditional heuristic-based tools in detecting complex smells like Refused Bequest, God Class, and Feature Envy.

In ~\cite{mesbah2025leveraging} authors evaluated LLMs for detecting code smells in the Machine Learning Code Quality (MLCQ) dataset ~\cite{madeyski2020mlcq}. Their study comparing GPT-4 and Llama found that prompt tuning significantly enhances LLM performance in detecting and classifying code smells. However, challenges remain in ensuring consistent detection across different smell types, as some LLMs struggle with distinguishing nuanced smells, leading to false positives and inconsistencies.

In the context of AI-powered code reviews, LLMs have been integrated into automated review systems to detect code smells and recommend improvements in real-time ~\cite{rasheed2024ai}. These systems leverage LLMs' ability to predict future code issues, identify design flaws, and enhance software maintainability. Similarly, research on LLMs for test smell detection has shown promising results, with models like ChatGPT-4 identifying 70\% of test smell types across multiple programming languages~\cite{lucas2024evaluating}.

\section{LLM-Generated Malicious Code}\label{intro}
Recently, LLMs have been utilized to generate code for users. It facilitates generating code effortlessly and saves time. However, it has increased the risk of cyber threats to the community. The rising use of LLMs for code generation brings into question their ability to produce reliable and trustworthy code ~\cite{mousavi2024investigation}. Attackers can use LLMs to generate malicious code automatically, endangering many software service providers (SSPs).
LLMs are akin to a double-edged sword: in the hands of a bodyguard, they can offer protection, but if wielded by an adversary, they can cause harm ~\cite{ALTORFER2025}. This situation calls for an investigation to understand the dynamics of malicious code generation by LLMs and detect the threats effectively. Notably, the existing academic research on malware generation using LLMs remains scarce, with only a limited number of studies addressing this issue ~\cite{botacin2023gpthreats}. It is intriguing to explore how existing techniques address the challenges posed by LLMs in restricting malware generation. 
Several researchers have observed that LLMs do not generate malicious code solely based on the direct prompt given that asks to create one ~\cite{botacin2023gpthreats, pa2023attacker}. LLMs pose several limitations to restrict the generation of malicious content as discussed in the subsection below: 

\subsection{Obstacles in Generating Malware using LLMs}
Regardless of the ability of LLMs to generate code, there are several hurdles and challenges in using them to synthesize fully functional malware. These obstacles arise due to ethical safeguards, AI limitations, and technical constraints. Some of the obstacles are listed below:

\begin{itemize}
    \item LLMs incorporate built-in safety mechanisms to prevent the generation of malicious code ~\cite{pa2023attacker}. If a prompt is identified as harmful, the model either refuses to respond or generates a generic or incomplete answer.
    \item Jailbreak prompts as discussed in subsection~\ref{jailbreak} can occasionally circumvent safeguards; however, models are continuously updated to prevent such exploits~\cite{shen2024anything}. OpenAI and other developers actively track and refine restrictions to block prompts requesting malware, hacking tools, or unauthorized scripts.
    \item LLMs generate code based on patterns learned from data but lack a deep understanding of complex exploit development, privilege escalation, or advanced evasion techniques. This absence of contextual awareness limits their ability to create sophisticated malware autonomously~\cite{haldar2024analyzing}.
    \item LLMs can explain program execution flow for simple programs, their performance declines with increasing code complexity, particularly when dealing with non-trivial logical and arithmetic operations, non-primitive types, and API calls~\cite{liu2024codemind}.
    \item Developing and deploying malware is illegal under various cybercrime laws, and using AI for such purposes raises significant ethical concerns~\cite{gupta2023chatgpt}. The potential for AI-generated content to be used maliciously has been a topic of discussion among experts and legislators, emphasizing the need for responsible AI development and regulation~\cite{alahmed2024exploring}.
\end{itemize}

\subsection{Techniques Applied for Malware Generation}
LLMs are highly speculated to cause a cybersecurity threat ~\cite{pa2023attacker}.  Figure~\ref{fig:mal_code_gen} shows the latest techniques used in literature to generate malicious code. Additionally, Table 4 provides an overview of research publications that explore the use of LLMs for malware code generation. It categorizes studies based on the techniques used, the LLM models involved, the datasets referenced, the intended applications of generated malware, and the effectiveness of security detection mechanisms against such threats.
Let's discuss the malware generation techniques in the Sub-sections below:
\begin{figure}
    \centering
    \includegraphics[width=0.45\textwidth]{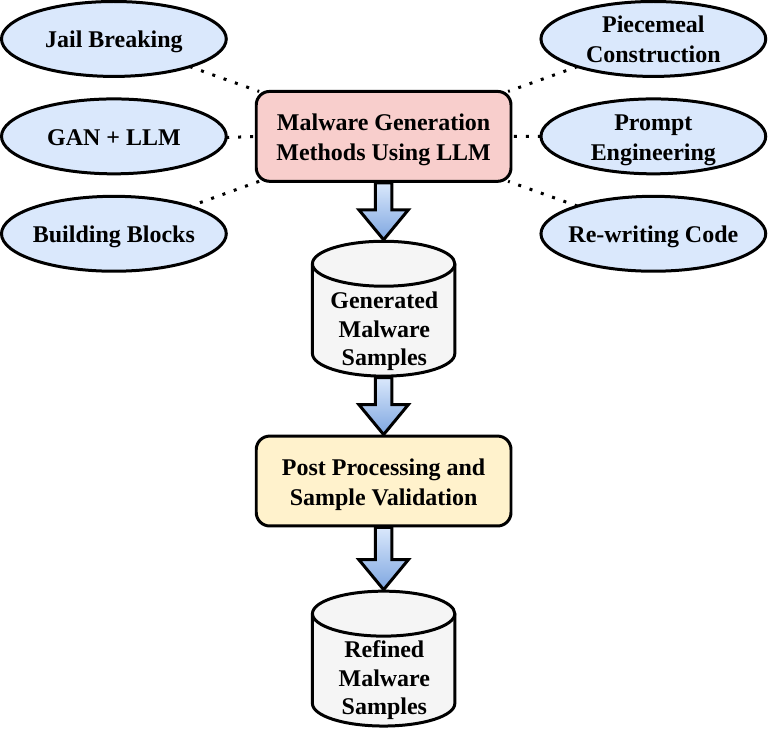}
    \caption{A general view of LLM-based malware generation techniques to generate malware sample datasets.}
    \label{fig:mal_code_gen}
\end{figure}

\subsubsection{Prompt Engineering to Bypass AI Constraints}
As mentioned above, LLMs impose safety and ethical restrictions on generating malicious code. Charan et al.~\cite{charan2023text} proposed prompt engineering techniques to bypass these safeguards and generate malware capable of concealing PowerShell scripts while executing them on a predefined schedule. This approach suggests employing LLMs to generate payloads for executing cyber attacks. The study demonstrates the high efficiency of LLMs by producing executable code for the top 10 MITRE weaknesses identified in 2022 using ChatGPT~\cite{ChatGPT} and Bard~\cite{GoogleBard}. Furthermore, LLM-generated payloads are often more sophisticated and precisely targeted compared to those crafted manually.

\subsubsection{Malware Generation using Building Blocks}
The building blocks approach to malware generation explored in the article~\cite{botacin2023gpthreats} presents a modular methodology where LLMs like GPT-3 generate malware in small functional segments rather than as a single, complete piece of code. This approach is highly effective because while GPT-3 struggles to create entire malware from a single prompt, it excels at producing individual malicious functions when requested separately. The attackers can break down complex malware functionality into discrete building blocks such as file manipulation, process injection, privilege escalation, and persistence mechanisms. These functional components can then be manually or programmatically integrated into a fully operational malware sample. 

Furthermore, GPT-3’s ability to generate multiple implementations of the same behavior using different APIs helps evade signature-based detection systems. The paper demonstrates this approach by generating thousands of malware variants, some of which were detected by as few as 4 out of 55 antivirus engines on VirusTotal. This highlights the dual-use nature of LLMs, where their text-generation capabilities can be exploited to automate and scale malware development, reducing the skill barrier for cybercriminals.

\subsubsection{Malware generation using JailBreaking}\label{jailbreak}

Jailbreaking is one of the recent strategies that allows the DAN (Do Anything Now) prompts to bar the restrictions of generative models such as ChatGPT~\cite{jailbreakCGpt, Lee_ChatGPT_DAN_2023}. More specifically, prompt engineering is employed to jailbreak ChatGPT~\cite{white2023prompt}.  Many researchers have explored jailbreaking attacks to exploit LLMs, such as~\cite{li2023multi, huang2023catastrophic}. 

Pa et al.~\cite{pa2023attacker} utilized jailbreaking prompts to bypass moderation controls and successfully generate functional malware and attack tools. The authors confirm that ChatGPT~\cite{ChatGPT}, text-davinci-003~\cite{textDavinci}, and Auto-GPT~\cite{AutoGPT} can be exploited using jailbreaking prompts to bypass safety restrictions, iterative prompt refinement to improve malicious code generation, and automated multi-step prompt generation using Auto-GPT. The researchers successfully developed seven types of malware and two attack tools, including ransomware, worms, keyloggers, fileless malware, and AI-powered phishing mailers. The study highlights that jailbreaking ChatGPT with specific prompts significantly increases the likelihood of generating malicious code, while Auto-GPT autonomously refines its prompts, making it a powerful tool for malware development. 
The study found that malware generated using GPT-3.5-turbo exhibited low detection rates across various antivirus (AV) and endpoint detection and response (EDR) solutions. Additionally, Linux-based malware, including IoT worms and Telnet brute-force tools, had detection rates below 3\%, making them particularly effective at bypassing traditional security measures. These results highlight the significant evasion capabilities of LLM-generated malware and the urgent need for advanced security strategies to mitigate AI-driven attacks.
Despite this, the model’s outputs contained frequent errors, requiring manual correction to execute attacks successfully.

\subsubsection{Piecemeal Construction}
The concept of piecemeal construction in malware generation, as demonstrated by Monje et al.~\cite{monje2023being} in their study, highlighted a method of bypassing AI guardrails by requesting individual benign code snippets that, when combined, form a fully functional malware program. Instead of directly asking ChatGPT to generate a ransomware script - an action that would trigger OpenAI’s ethical safeguards - the authors successfully circumvented restrictions by breaking down their requests into multiple, seemingly harmless programming tasks. These included generating encryption and decryption functions, creating a pop-up window for ransom demands, and monitoring cryptocurrency payments. The researchers constructed a functional ransomware program by stitching together these individual scripts, demonstrating a significant security loophole in LLM-based AI systems. This piecemeal approach not only exposes the vulnerabilities of LLMs in content moderation but also raises concerns about the misuse of generative AI for malicious cyber activities.

\subsubsection{Re-writing Code for generating Malware Variants}
Regarding code generation using various techniques, some researchers have explored the potential of LLMs in generating LLM-assisted malware variants. These variants can be created by rewriting the original source code of malware samples using LLMs.

In this direction, Huang et al.~\cite{huang2024exploring} investigated the effectiveness of LLMs, specifically OpenAI’s GPT-3.5-Turbo, in generating malware variants and their ability to evade detection by antivirus software. The researchers developed 400 malware variants from four different types of malware (worm, keylogger, ransomware, and fileless malware) by modifying their source code using GPT-3.5-Turbo.

For each type of malware, 100 variants were generated, 211 of them compiled into fully functional executables. The similarity between the original and generated malware was assessed using BLEU scores and binary similarity metrics. The study also examines antivirus detection methods, evaluating six antivirus solutions using 160 malware variants. Detection techniques included both file-based scanning and behavior monitoring. The results indicate that fileless malware had the lowest detection rate, while ransomware was more likely to be identified through behavioral analysis. The worms and keyloggers exhibited varying evasion rates for each antivirus software, with some solutions not detecting certain variants.

Similarly, Botacin et al.~\cite{botacin2023gpthreats} suggested that GPT-3 can produce different versions of the same malware by rewriting existing malicious code in multiple ways. The study generated 4,820 functional malware variants, demonstrating GPT-3’s ability to produce diverse and evasive malware samples.
The detection rates of these variants varied significantly, and some samples evaded detection by most antivirus engines.

\subsubsection{GAN and LLM for Malware Generation}
Researchers have adapted LLMs and other generative AI models for various cybersecurity applications. For example, Devadiga et al. ~\cite{devadiga2023gleam} proposed a GLEAM model (GAN and LLM for Evasive Adversarial Malware) utilizing GPT-3 for the generation of synthetic malware, particularly in the creation of hexadecimal representations of malware samples. LLM-generated malware was tested against machine learning-based malware detectors, and the results showed that GPT-3-generated malware had a higher evasion rate than GAN-generated samples. The study demonstrated an increase in the evasion rate when GPT-3-generated samples were used, highlighting its potential threat in adversarial malware generation.

\begin{table*}[ht!]\label{tab4LLMs}
\caption{Publications corresponding to LLMs used for malware code generation.}
\begin{tabular}{ p{1.8cm} p{1.5cm} p{2cm} p{1.5cm} p{1.5cm} p{2cm} p{3cm} }
\hline
\textbf{Paper/Year}                         & \textbf{Techniques used}    & \textbf{Mitre Tactics}                            & \textbf{LLM model(s)}                       & \textbf{Dataset Used}                                      & \textbf{Application}                       & \textbf{Security Detection}          \\ \hline
Pa et al. 2023~\cite{pa2023attacker}           & Jailbreaking       & Initial access-Execution-Defence Evasion & ChatGPT, Text-Davinci-003, AutoGPT & N/A                        & Generate malware and attack tools & Partially detected (low detection rates)             \\ \hline
Botacin et al. 2023~\cite{botacin2023gpthreats}   & Building Blocks    & Execution                                & GPT-3                              & Malware samples~\cite{botacin2020malwaresample} & Code generation and DLL injection & Some codes are detected by antiviruses; some are not \\ \hline
Charan et al. 2023~\cite{charan2023text}           & Prompt Engineering & Top 10 Mitre Techniques from Red Report 2023~\cite{redReport2023}                               & ChatGPT, Google Bard               & The Red Report 2023~\cite{redReport2023}     & Malicious Code Generation         & Detection evasion in controlled sandbox environment                                                  \\ \hline
Monje et al. 2023~\cite{monje2023being}    & Piecemeal Construction    & Execution                                & GPT-3                              & System files & Ransomware generation & Some codes are detected by antiviruses; some are not \\ \hline

 Liu et al. 2023~\cite{liu2023jailbreaking}    & Jailbreaking    & Execution                                & GPT-3                              & Custom-built scenario
\& jailbreak prompt datasets & Identification of effective jailbreak prompts for malware generation & Some codes are detected by antiviruses; some are not \\ \hline
 
 Huang et al. 2024~\cite{huang2024exploring}   & Prompting LLM to re-write code    & Intial access-Execution-Defence Evasion-Credential Access                                & GPT-3.5-Turbo            & Malware samples (source unspecified) & Malware Variants Generation & Partially effective at evading antivirus detection\\ \hline

Devadiga et al. 2023~\cite{devadiga2023gleam}    & GAN based attacks    & Execution-Defence Evasion-Credential Access                                & GPT-3            & PE Malware Machine Learning Dataset~\cite{PEDataset2021}  & Evasive Malware Samples Generation & Partially effective at evading antivirus detection\\ \hline

\end{tabular}
\end{table*}

\section{Top LLM-Based Models for Malware Analysis}
\begin{figure*}[h]
    \centering
    \includegraphics[width=0.90\textwidth]{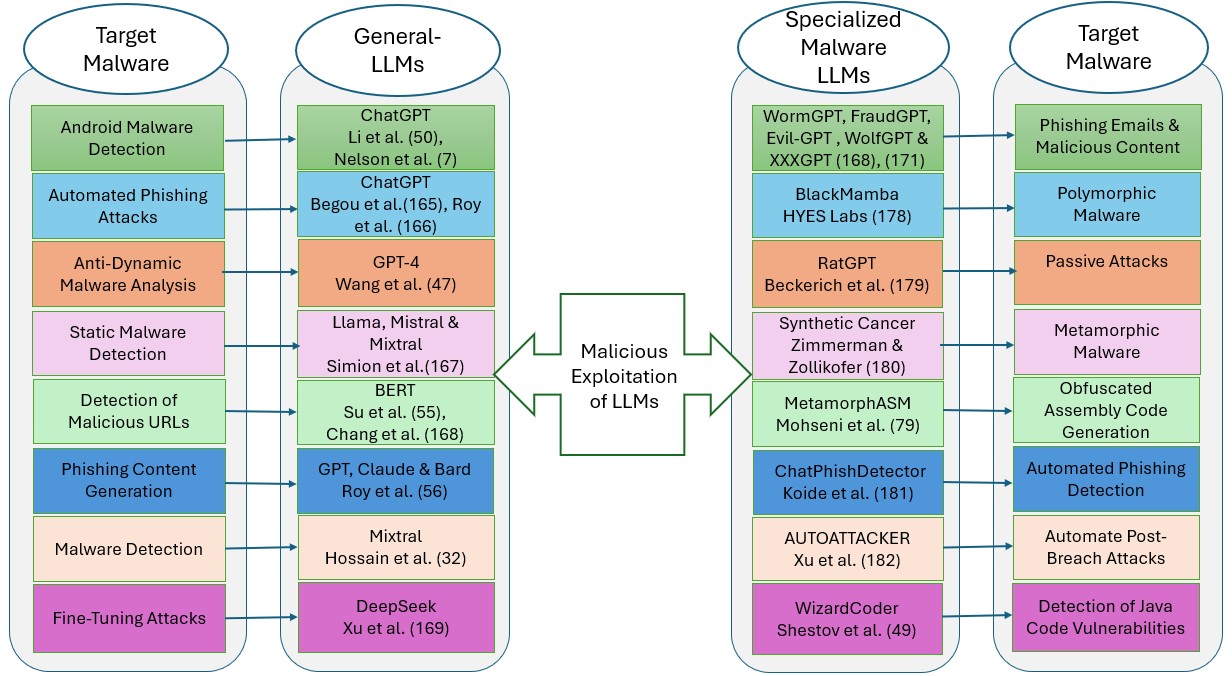}
    \caption{Application of different LLMs for malware code analysis across various targets.}
    \label{MalwareLLMs}
\end{figure*}

In this section, we explore LLMs that demonstrate exceptional capabilities in malware code analysis. These models can be categorized into two main groups. The first group includes general-purpose LLMs, which are designed for a wide range of tasks and exhibit strong natural language and code processing abilities. While not specifically trained for malware detection, these models can be adapted for malware analysis by leveraging their foundational knowledge of programming languages and code structures. The second group consists of specialized offensive AI models that are specifically developed or fine-tuned for malicious purposes such as generating phishing attacks, obfuscating code, or automating cyberattacks, enabling them to detect, analyze, and classify malicious code with greater accuracy. These specialized models are particularly valuable for cybersecurity research, as they enhance the ability to identify evasive threats and emerging attack patterns.
Figure~\ref{MalwareLLMs} illustrates the malicious exploitation of LLMs, categorizing them into general-purpose LLMs and specialized malware LLMs, along with their applications in cybersecurity threats. It maps how various LLMs contribute to different malware-related tasks, including phishing, malware detection evasion, code obfuscation, and automated cyberattacks.

\subsection{Maliciously Exploited General-Purpose LLMs}
Maliciously Exploited LLMs refer to the misuse of advanced AI-driven text generation systems for cybercrime, misinformation, and automated attacks. While GPT-4 \cite{GPT-4}, DeepSeek ~\cite{guo2024deepseek}, Claude \cite{claude2023}, and Gemini are designed for productive and ethical applications, they can be manipulated to generate harmful content, including phishing emails, malware code, misinformation campaigns, and social engineering scripts. Cybercriminals exploit vulnerabilities in LLMs by bypassing safety filters using adversarial prompts, jailbreak techniques, and fine-tuning attacks, enabling models to produce highly deceptive and malicious responses. A growing concern is the use of LLMs in malware development, where models assist in code obfuscation, polymorphic malware generation, and automated exploit scripting, significantly lowering the skill barrier for cybercriminals. Additionally, LLM-driven social engineering enables attackers to craft personalized phishing messages and fraudulent interactions, making them more convincing and harder to detect. The exploitation of LLMs poses serious security risks, necessitating stronger AI guardrails, real-time monitoring, and regulatory frameworks to mitigate their misuse in cybersecurity threats.

\subsubsection{OpenAI models}
The advent of OpenAI models has profoundly impacted the field of artificial intelligence. Researchers are actively exploring the latest GPT (Generative Pre-trained Transformer) series, including GPT-3, GPT-3.5 Turbo, GPT-4o~\cite{GPT-4}, GPT-4mini~\cite{GPT-4mini}, and ChatGPT for malware generation, automated phishing attacks, malware detection, and analysis. 

Li et al.~\cite{li2024enhancing} examined how ChatGPT enhances android malware detection by improving interpretability rather than making direct classification decisions. Traditional models like Drebin~\cite{arp2014drebin}, MaMaDroid~\cite{onwuzurike2019mamadroid}, and $XM_{AL}$ ~\cite{wu2021android} are effective but suffer from bias and lack of transparency, making it difficult for developers to trust their outputs. The study integrates ChatGPT into malware analysis workflows, using static and dynamic analysis techniques to extract app features and employing prompt engineering to generate detailed explanations of potential threats. While ChatGPT does not classify malware, it helps security analysts understand detection results, improving usability, trust, and decision-making in malware research. The findings highlight ChatGPT’s value as an interpretability tool, complementing traditional malware detection systems by offering comprehensive explanations of security threats.

In addition to malware generation and detection, the GPT models automate phishing attacks. For example, Begou et al.~\cite{begou2023exploringdark} examined how ChatGPT, based on the GPT-3.5-turbo-16K model, can be exploited to develop and automate sophisticated phishing attacks. The study demonstrates how ChatGPT can assist in various aspects of phishing, including website cloning, credential theft integration, code obfuscation, domain registration, and phishing site deployment. The research highlights the vulnerabilities of AI-driven text models, revealing that despite safety measures, ChatGPT can be manipulated to generate and automate phishing websites that closely mimic legitimate ones. \newline
Similarly, in another study, Roy et al.~\cite{roy2023generating} demonstrated how ChatGPT can be misused to generate sophisticated phishing emails, highlighting the potential for AI-driven tools to enhance the effectiveness of social engineering attacks.
 
Wang et al. ~\cite{wang2024unmasking} proposed an LLM-based workflow to analyze and detect anti-dynamic analysis techniques (TADA) in malware, which are designed to evade sandbox detection, debugging, and reverse engineering efforts. The authors incorporated the GPT-4 Turbo model to analyze the malware's assembly code features, string-based indicators, and API call sequences. GPT-4 was able to extract API call patterns, analyze assembly code features, and detect obfuscated strings used for sandbox evasion. 

\subsubsection{Llama}\label{llama}
Llama 2, developed by Meta, is an open-weight Large Language Model (LLM) optimized for text generation, reasoning, and coding tasks. Due to its availability and ease of fine-tuning, it has been leveraged for cybersecurity research but also misused for malware generation, phishing automation, and exploit development by cybercriminals who modify its training data. Its open-source nature makes it a powerful yet dual-use AI system, requiring strict governance to prevent adversarial applications.

Simion et al.~\cite{simion2024benchmarking} evaluated the efficacy of open-source LLMs in detecting malware by analyzing sequences of API calls. The researchers assessed four LLMs—Llama2-13B, Mistral, Mixtral, and Mixtral-FP16—using a dataset comprising 20,000 malware samples and 20,000 benign files. The study concluded that when used without fine-tuning, these models exhibited low accuracy and were unsuitable for real-time malware detection scenarios. This underscores the necessity for fine-tuning LLMs and integrating them with traditional security tools to enhance their effectiveness in identifying malicious software.

\subsubsection{Mixtral}
Mixtral is a Mixture-of-Experts (MoE) model developed by Mistral AI, combining multiple expert models to improve scalability and accuracy while maintaining lower computational costs.
Hossain et al.~\cite{hossain2024malicious} suggested an approach to detect malicious code in Java source files, incorporating the Mixtral model. Traditional signature-based and behavioral-based malware detection techniques struggle with zero-day attacks and obfuscation techniques, prompting the need for AI-driven solutions. The research introduces an LLM-powered detection system that analyzes Java code structures using graph-based methodologies.

\subsubsection{BERT} BERT (Bidirectional Encoder Representations from Transformers): Created by Google, BERT is designed to understand the context of a word in search queries, improving the accuracy of search results.\newline

Su et al.~\cite{su2023bert} inspected BERT's transformer-based architecture to enhance the detection of malicious URLs by capturing semantic relationships and contextual patterns in URL structures. Unlike traditional approaches that rely on lexical features or shallow machine learning models, BERT processes URLs using tokenization, self-attention, and deep contextual embeddings, allowing it to distinguish benign, phishing, malware, and defacement URLs with higher accuracy.  Additionally, BERT demonstrates versatility by analyzing both raw URLs and structured feature representations (e.g., extracted API calls and request patterns) via multi-view input processing, making it applicable to various cybersecurity domains, including phishing detection, DNS-over-HTTPS attack identification, and IoT malware analysis. The results confirm BERT’s performance in malicious URL detection, proving its effectiveness in real-time cybersecurity applications.

Similarly, in another study, Chang et al.~\cite{chang2021research} proposed a method that integrates the BERT model with Convolutional Neural Networks (CNN) to enhance malicious URL detection. This approach effectively captures both the contextual relevance and local features of URLs, leading to improved detection accuracy.

\subsubsection{Claude}\label{claude}
Claude~\cite{claude2023}, developed by Anthropic, is an AI-powered conversational model designed with a focus on safety, interpretability, and ethical AI alignment. Bard~\cite{Bard}, rebranded as Gemini by Google DeepMind, is an advanced AI model designed for conversational AI, research assistance, and code generation.
Roy et al. ~\cite{roy2024chatbots} inspected how commercially available LLMs, specifically ChatGPT (GPT-3.5 Turbo), GPT-4, Claude, and, Bard, can be manipulated to generate phishing websites and emails. The study reveals that these LLMs can circumvent content moderation and produce highly convincing phishing attacks that mimic legitimate brands using recursive prompt engineering techniques. Attackers can automate and scale phishing content generation by reusing LLM-generated prompts, reducing the manual efforts required. Additionally, the authors disclose their findings to major LLM providers, leading to security updates. The research highlights the dual-use nature of LLMs, underscoring both their potential for malicious exploitation and the development of countermeasures to mitigate risks.

\subsubsection{DeepSeek}
DeepSeek~\cite{bi2024deepseek} is an open-weight LLM developed for advanced reasoning, code generation, and multilingual processing. Its high adaptability and fine-tuning capabilities have made it a valuable tool for cybersecurity research, including malware detection, exploit analysis, and penetration testing. The paper~\cite{xu2025dark} investigates the security risks associated with DeepSeek-R1, a Chain-of-Thought (CoT) reasoning-based LLM. The study specifically examines fine-tuning attacks, which manipulate the model’s output by bypassing its safety alignment and ethical safeguards. The researchers conduct adversarial fine-tuning on DeepSeek-R1-Distill-Llama-8B, a distilled version of DeepSeek, using a dataset of harmful prompts. Experimental results reveal that fine-tuned DeepSeek exhibits a significant increase in generating harmful content, highlighting its vulnerability to malicious modifications. Compared to non-CoT models like Mistral-7B, DeepSeek’s step-by-step reasoning process amplifies the generation of dangerous responses, making it particularly susceptible to security risks. The study underscores the critical need for stronger safeguards, regulatory oversight, and robust security mechanisms to prevent the exploitation of CoT-enabled LLMs for unethical or malicious purposes.

\subsection{Specialized Offensive AI Models and Applications}
Unlike general-purpose LLMs, these models are specifically designed or fine-tuned for malicious purposes. They are developed without imposing any ethical restrictions and are used for phishing automation, malware obfuscation, exploit creation, and social engineering. These models are commonly found on dark web marketplaces and underground hacking forums, providing cybercriminals with advanced AI-driven tools. Furthermore, there are various malicious applications of LLMs are also discussed in the following subsections:

\subsubsection{WormGPT}
WormGPT~\cite{WormGPT2023} is a malicious LLM developed by cyber criminals to facilitate fraudulent activities. Unlike legitimate LLMs designed for beneficial applications, WormGPT is tailored to assist in creating sophisticated phishing emails, crafting malicious code, and automating various cyberattacks~\cite{falade2023decoding}. Its primary usage revolves around enabling cybercriminals to execute more convincing and efficient fraudulent schemes utilizing advanced language generation capabilities. 

\subsubsection{FraudGPT}
FraudGPT is an advanced AI-driven tool that has been specifically designed to assist cybercriminals in conducting fraudulent activities at scale. FraudGPT is programmed without safety measures, making it a powerful tool for launching cyberattacks, including credential theft, phishing, malware development, and scam automation~\cite{falade2023decoding}. It enables malicious actors to generate convincing phishing emails, social engineering scripts, and exploit kits that evade traditional cybersecurity defences~\cite{fraudGPT2023}.

\subsubsection{Evil-GPT}
Evil-GPT is a malicious AI chatbot that emerged on darknet forums, designed explicitly for cybercriminal activities~\cite{evilGPT-2}. Marketed as an alternative to tools like WormGPT, Evil-GPT is built on open-source LLMs and fine-tuned to help in generating malicious code, crafting sophisticated phishing emails, and developing hacking tools~\cite{evilGPT}. Its availability on underground markets lowers the barrier to entry for cybercriminals, enabling individuals with limited technical expertise to conduct complex cyberattacks.

\subsubsection{DarkGPT}
DarkGPT \cite{DarkGPT} represents a new category of LLMs specifically designed to function without conventional ethical constraints or content restrictions. Essentially, it serves as an "unfiltered" variant of ChatGPT, capable of generating illicit or sensitive human-like text \cite{DarkGPT_Explore}. It has been identified as one of several malicious AI chatbots marketed to cybercriminals, bypassing safety checks via "jailbreaking" to generate content that mainstream models would normally refuse~\cite{aiid:736}.
DarkGPT is an advanced AI model capable of generating highly malicious code, excelling in developing sophisticated malware that not only compiles successfully but also evades antivirus detection. ~\cite{lin2024malla}.

\subsubsection{Wolf GPT}
Wolf GPT is a malicious AI chatbot that has emerged on dark web forums. It is designed to assist cybercriminals in their illicit activities. Wolf GPT is built using open-source LLMs, it enables users to generate malicious code, craft sophisticated phishing emails, and develop hacking tools. Its availability lowers the barrier to entry for cybercriminals, allowing individuals with limited technical expertise to conduct complex cyberattacks.  

\subsubsection{XXX GPT}
XXX GPT is another malicious AI chatbot that has been identified on hacker forums, offering cybercriminals a platform to automate and enhance their malicious endeavors. Similar to Wolf GPT, XXX GPT leverages advanced AI capabilities to facilitate the creation of malware, phishing schemes, and other cyber threats~\cite{blackGPT}. 

\subsubsection{BlackMamba} 
BlackMamba is a proof-of-concept malware developed by HYAS Labs~\cite{BlackMamba} that leverages artificial intelligence to dynamically generate polymorphic keylogging code at runtime, effectively evading traditional endpoint detection and response (EDR) systems. Upon execution, BlackMamba reaches out to OpenAI's API to synthesize malicious code, which is then executed within the benign program's context using Python's 'exec()' function, leaving no trace on the disk. This process allows BlackMamba to re-synthesize its keylogging capability with each execution, rendering the malicious component truly polymorphic. Captured keystrokes, including sensitive information such as usernames, passwords, and credit card numbers, are exfiltrated via legitimate communication channels like Microsoft Teams, further complicating detection efforts. 

\subsubsection{RatGPT}
Beckerich et al.~\cite{beckerich2023ratgpt} explore how LLMs, particularly ChatGPT, can be misused as intermediaries for malware attacks. The study demonstrates a proof-of-concept attack where an LLM is used as a proxy between an attacker and a victim, bypassing traditional cybersecurity measures. The researchers show how ChatGPT-generated payloads can be used to weaponize seemingly harmless executables, enabling them to establish communication with a Command and Control (C2) server. By exploiting web-accessible plugins, the malware can dynamically generate IP addresses, execute commands, and exfiltrate data while remaining undetected.  
A key finding is that ChatGPT and similar LLMs can facilitate passive attacks, allowing malware to operate without direct interaction from the victim. The attack model relies on LLM-generated code for both payload delivery and remote execution, circumventing traditional malware detection methods since no hardcoded malicious code is stored on disk. 

\subsubsection{Synthetic Cancer}
Zimmerman and Zollikofer ~\cite{zimmerman2024synthetic} present a prototype of metamorphic malware that utilizes LLMs for automatic code rewriting to evade signature-based detection. LLM alters malware's source code during replication, generating unique variants with each execution. This constant evolution allows it to bypass signature-based detection mechanisms employed by traditional antivirus programs. Additionally, the malware employs social engineering techniques, analyzing previous email conversations to generate contextually relevant and persuasive responses. These responses include infected attachments, tricking recipients into opening them, and inadvertently spreading the malware further.

\subsubsection{MetamorphASM}
Mohseni et al.~\cite{mohseni2024can} examine how LLMs can generate obfuscated assembly code, challenging antivirus systems. They introduce the MetamorphASM benchmark with 328,200 obfuscated code samples, using techniques like dead code insertion and register substitution. The study evaluates LLMs like GPT-3.5, GPT-4, and others, using information-theoretic metrics and human reviews for assessment

\subsubsection{ChatPhishDetector} 
Koide et al. ~\cite{koide2024chatphishdetector}, introduced ChatPhishDetector, a novel system that uses LLM to identify phishing websites. The system employs a web crawler to gather comprehensive website data, which is then transformed into prompts for LLM analysis. This approach enables the detection of multilingual phishing sites by recognizing impersonated brands and social engineering tactics without the need for additional machine learning training.

\subsubsection{AUTOATTACKER} The study~\cite{xu2024autoattacker} explored the performance of GPT-3.5-Turbo and GPT-4 within the AUTOATTACKER framework to automate post-breach cyberattacks such as privilege escalation, credential theft, and ransomware execution.  It is an innovative LLM-based system to automate complex, human-like cyber-attacks, commonly known as "hands-on-keyboard" attacks. The system comprises four key components—Summarizer, Planner, Navigator, and Experience Manager that work together to generate precise attack commands tailored to various simulated environments. Extensive testing revealed that while models like GPT-3.5 and Llama2 variants showed limited effectiveness, GPT-4 demonstrated remarkable capabilities in conducting post-breach attacks with minimal human intervention.


\subsubsection{WizardCoder}
Shestov et al.~\cite{shestov2024finetuning} proposed WizardCoder, an advanced variant of StarCoder, which plays a crucial role in detecting vulnerabilities in Java source code. The model, with its 13-billion parameter architecture, is fine-tuned specifically for binary classification of vulnerable vs. non-vulnerable code functions. Compared to CodeBERT-based models, WizardCoder demonstrates higher accuracy and robustness in identifying security flaws. The paper highlights key improvements, such as adapting WizardCoder for classification tasks, handling imbalanced datasets, and optimizing training speed using batch-packing techniques. The fine-tuned WizardCoder outperforms ContraBERT, achieving better ROC AUC and F1 scores on both balanced and imbalanced datasets. 

\section{Malware Datasets}
Researchers need comprehensive and diverse malware datasets to analyze malware techniques and attack strategies effectively. One common approach to obtaining such datasets is through online platforms associated with anti-malware projects, including MalShare, Malware DB, and VirusShare. In addition, some organizations and research groups periodically release malware data sets to support cybersecurity research and mitigate the scarcity of publicly available sources. 

Beyond direct collection, researchers have begun to explore the generation of synthetic malware datasets as an alternative. In particular, LLMs have emerged as a promising tool for generating malware code datasets, using the techniques discussed earlier in Section \ref{intro}. LLMs can create synthetic malware samples, obfuscate malicious code, and generate variations of known threats, effectively expanding and diversifying existing malware datasets. Furthermore, data augmentation has become a powerful technique in enhancing malware detection models. For example, LLMs can simulate polymorphic and metamorphic malware, allowing machine learning-based detection systems to improve accuracy and robustness by training on adversarially generated malware variants. This approach has been instrumental in strengthening threat intelligence databases, improving malware classifiers, and enhancing cybersecurity defences against rapidly evolving threats.

\subsection{Ready-to-use Malware Datasets}
Existing Malware datasets are essential in cybersecurity research, providing structured collections of malicious software samples to aid in detection, analysis, and prevention. These datasets cover a range of malware types and attack vectors, including Android malware (AndroZoo, DREBIN, KronoDroid) and Windows PE malware (BODMAS, SOREL-20M).  

Additionally, repositories like VirusTotal and MalwareBazaar provide continuous updates of real-world malware samples, ensuring researchers have access to evolving threats. Public (open) datasets are freely available or have minimal restrictions, enabling data-driven research, AI model training, and standardized benchmarking for malware detection. Restricted datasets require special permissions, passwords, or commercial access, offering high-quality, proprietary threat intelligence while maintaining security controls to prevent misuse. These datasets are crucial for identifying emerging threats, enhancing AI-driven security models, and enabling reliable cybersecurity solutions. Table 5 lists a mix of open and restricted malware data resources that can be used to obtain malware samples. Furthermore, a concise overview of some of the prominent datasets is provided below:
\begin{itemize}
    \item \textbf{AndroZoo~\cite{AndroZoo2016Allix}} A comprehensive repository of Android malware samples, widely used for machine learning-based malware classification.
    \item \textbf{BODMAS~\cite{yang2021bodmas}} A data set containing time-stamped malware samples and well-curated family information for research purposes.

    \item \textbf{DikeDataset \cite{Dike_link}} This dataset comprises labeled samples, including benign and malicious portable executable (PE) and Object Linking and Embedding (OLE) files. 
    \item \textbf{DREBIN~\cite{arp2014drebin}} A large-scale dataset for static and dynamic analysis of Android malware.
    \item \textbf{EMBER~\cite{anderson2018ember}} A dataset for portable executable (PE) malware analysis, widely used in machine learning-based malware feature engineering.
    \item \textbf{MalRadar~\cite{wang2022malradar}} A data set that focuses on Android malware, providing disarmed samples for research purposes. 
    \item \textbf{Malware Datasets Top-1000 Imports~\cite{mal1000dataset_link}}  It is a static malware analysis dataset that contains the top 1,000 imported functions from Windows PE files.
    \item \textbf{MalwareBazaar~\cite{MalwareBazaar2024}} A public threat intelligence sharing platform that offers access to diverse zero-day malware samples.
    \item \textbf{Malware Exploratory~\cite{MalwareExploratory_link}} This dataset provides an in-depth analysis of malware samples, offering insights into malware behavior, patterns, and classifications to support cybersecurity research and threat intelligence.
    \item \textbf{Microsoft Malware Classification Challenge ~\cite{ronen2018microsoft}} A dataset providing labeled Windows executable malware samples, often used for deep learning-based malware detection.
    \item \textbf{MOTIF ~\cite{joyce2023motif}} It is the largest public malware dataset with ground truth family labels, facilitating malware classification and behavior analysis research.
    \item \textbf{SOREL-20M ~\cite{harang2020sorel}} This dataset comprises metadata and features for approximately 20 million Windows Portable Executable (PE) files, including 10 million disarmed malware samples. 
    \item \textbf{TheZoo~\cite{theZoo}} This is a live malware repository that is maintained for educational and research purposes.
    \item \textbf{VirusShare~\cite{VirusShare_link}} A well-known repository of multi-platform malware samples, but access is restricted to verified security researchers.
    \item \textbf{Vx-underground~\cite{VxUnderground}} It is one of the largest publicly available malware repositories, providing over 35 million malware samples, including source code and research materials.
\end{itemize}

\begin{table*}[!ht]
\caption{Some noteworthy applications of LLMs in malware analysis.}
    \centering
    \begin{tabular}{p{2.5cm} p{2.5cm} p{2.5cm} p{2.5cm} p{3cm} p{2.5cm}}
    \hline
        \textbf{Malware Dataset (link)} & \textbf{Application} & \textbf{Target} & \textbf{Access Status} & \textbf{Content} & \textbf{Number of Files} \\ \hline
        AndroZoo ~\cite{andro_link} & Malware Analysis & Android Malware & Requested Access & APK files, Metadata, VirusTotal Reports & 25M+ APKs \\ \hline
        BODMAS ~\cite{bodmas2022} & Malware Analysis & Windows PE Malware & Public Access & Labeled Windows PE malware dataset & 60M+ Samples \\ \hline
        Contagio Malware Dump~\cite{Contagio_link} & Malware Analysis & Various Malware Types & Password Protected & Clean and malicious files in various formats & 16,800 clean and 11,960 malicious files \\ \hline
        DikeDataset~\cite{Dike_link} & Malware Analysis & PE and OLE Files & Public Access & Labeled benign and malicious PE and OLE samples & Not specified \\ \hline
        DREBIN~\cite{Drebin_link} & Threat Detection & Mobile Threats & Requested Access & Android Malware Static Features & 5,560 Apps \\ \hline
        KronoDroid~\cite{krondroid_link} & Android Malware Analysis & Android Malware & Public Access & Hybrid dataset with static and dynamic malware features & Samples from 2008-2020 \\ \hline
        MalRadar ~\cite{malRadar_link} & Android Malware Analysis & Android Malware & Requested Access & Disarmed Android malware samples & 4,534 unique Android malware samples \\ \hline
        Top-1000 PE Imports~\cite{mal1000dataset_link} & Malware Analysis & Windows PE Imports & Registered Users & Top-1000 imported functions in PE files & Top-1000 Imports Dataset \\ \hline
        MalwareBazaar~\cite{MalwareBazaar2024} & Threat Intelligence & Zero-Day Malware & Public Access & Malware Samples and Signatures & Ongoing Updates \\ \hline
        Malware-Exploratory~\cite{MalwareExploratory_link} & Malware Analysis & General Malware & Open & Exploratory data analysis of malware samples & 216,352 Samples \\ \hline
        Microsoft Malware Classification Challenge~\cite{ronen2018microsoft}  & Malware Analysis & Windows Executables & Public Access & PE Files with Labels & 10M+ Samples \\ \hline
        MOTIF ~\cite{Motif_link} & Threat Detection & Malware Campaigns & Public Access & Attack Attribution and Threat Intelligence & 3,095 disarmed PE malware samples from 454 families \\ \hline
        SOREL-20M~\cite{Sorel20M_link} & Malware Analysis & Windows PE Malware & Public Access & Feature-extracted PE malware samples & 20M+ Samples \\ \hline
        TaintBench~\cite{TaintBench_link} & Android Taint Analysis & Real-world Android Malware & Public Access & Benchmark suite for evaluating Android taint analysis tools & 39 malware applications with 249 documented benchmark cases \\ \hline
        TheZoo~\cite{theZoo_link} & Malware Analysis & Multi-platform Malware & Public Access & Executable malware samples for research & Multiple Families of Malware \\ \hline
        VirusShare~\cite{VirusShare_link} & Malware Analysis & Multi-platform Malware & Registered Users & Malware samples across different platforms & Ongoing Updates \\ \hline
        Vx-underground~\cite{VxUnderground} & Malware Analysis & Multi-platform Malware & Public Access & Malware samples, source code, research papers & 35M+ Malware Samples \\ \hline
    \end{tabular}
\end{table*}

\subsection{Employing AI and LLMs for Malware Dataset Augmentation}
Data augmentation is a technique employed to artificially increase the size and diversity of a dataset by creating new samples through various transformations or modifications of existing data ~\cite{devries2017dataset}.
 Dataset augmentation plays a critical role in enhancing malware detection within the NPM ecosystem. Yu et al. ~\cite{yu2024maltracker} utilized LLMs to translate malicious code snippets from multiple programming languages (e.g., Python, C, Go) into JavaScript, significantly expanding the available dataset. This cross-language translation ensures that Maltracker captures a wider range of attack patterns, allowing it to detect malware that traditional rule-based methods might overlook. By integrating these synthetically generated yet functionally equivalent malware variants into the dataset, the model improves its ability to recognize polymorphic and obfuscated threats. This highlights how AI-powered dataset expansion can significantly enhance the robustness of malware tracking and cybersecurity defences in real-world software repositories like NPM.

Likewise, Palo Alto Networks Unit 42 researchers ~\cite{Palo42Alto} utilized  LLMs to iteratively transform malicious JavaScript code, generating numerous obfuscated variants that retained their original functionality but evaded detection by existing models. By integrating these LLM-generated samples into the training dataset, they effectively augmented the data, leading to improved model generalization. This approach demonstrates the efficacy of using LLMs for data augmentation to bolster the robustness of malware detection systems.

\section{Discussions, Limitations and Suggestions}
The field of LLMs for code analysis presents various challenges and open questions that warrant further research. Our study identifies key areas that remain insufficiently addressed and hold great potential for future advancements. In this section, we examine several critical challenges and identify gaps in existing work, outlining potential directions for future research.

\subsection{Future work}
The advancement of LLMs in malware analysis opens up numerous avenues for further exploration. In this section, we outline several potential directions for future work based on the current findings and limitations observed in our study. These proposed areas aim to address existing gaps, enhance the performance and reliability of LLM-based systems, and extend their applicability across diverse cybersecurity contexts. The following points highlight key research opportunities and practical developments that could shape the next generation of intelligent malware analysis tools.

\begin{itemize}
    \item{\textbf{Knowledge Graph Models with LLM for Malware:}}
    Knowledge graphs (KGs) are used to represent knowledge across different fields, helping to identify connections between entities. For example, by analyzing similarities between malware and attackers, KGs can reveal patterns. If multiple attackers use the same malware and techniques, they likely belong to the same organization. In a study by Hu et al. (2024) \cite{hu2024llm}, they proposed an automated method for building a threat intelligence knowledge graph. This method uses GPT's few-shot learning ability to generate prompts, allowing for automatic data annotation and augmentation. This reduces the need for manual work and creates datasets for training models. The Llama2-7B model was then fine-tuned to classify topics, extract entities and relationships, and identify TTPs (Tactics, Techniques, and Procedures) from threat intelligence reports, ultimately creating a useful knowledge graph.    
    
    \item{\textbf{Hybrid approaches for Malware Reverse Engineering:}}
     LLM models or non-LLM models can independently play a good role in reverse engineering on malware files, such as PE files. However, considering a hybrid approach based on LLM and reverse engineering tools such as IDA Pro or Ghidra can help disassemble or decompile files.
    
    \item{\textbf{Chain-of-Thought Prompt methods for decompiling:}}
    To understand the application of Chain-of-Thought Prompt methods for decompiling, we first need to consider the role of CoT prompting in complex tasks. CoT prompting involves breaking down a problem step-by-step, allowing the model to reason through each component before concluding. When applied to decompiling, this method enables a more structured and logical approach to reverse engineering code or text, making it easier to understand the underlying processes and components. However, despite its potential, it remains an open challenge. The complexity of decompiling code or intricate structures, combined with the limitations of current models, means that achieving fully reliable and efficient decompiling using CoT prompting is still a work in progress.
    
    For example, in \cite{fang2024stacksight}, the authors propose applying CoT prompting, a recent advancement in NLP that guides  LLMs through intermediate reasoning steps, to the task of decompiling WebAssembly binaries into high-level C code. This approach draws inspiration from human-like thinking processes to break down the complex task of decompilation into multiple phases, enabling the model to interpret, abstract, and translate binary code more effectively.
    
    \item{\textbf{LLM for Malware Code and DFA:}}
    DFAs are helpful in malware analysis to track data flow and spot suspicious behavior in source code. Combining DFAs with NLP models like LLMs can improve malware analysis by using the strengths of both. DFAs track data flow and dependencies, while LLMs can find patterns, malicious intent, or hidden threats in text. There is little work on these methods; for example, in \cite{wang2024llmdfa}, the authors created a customizable framework for dataflow analysis using LLMs to analyze Java programs. They used the tree-sitter library to extract key information like parameters, return values, callers/callees, and sources/sinks. The framework was tested with four LLMs: GPT-3.5, GPT-4, Gemini-1.0, and Claude-3, and its performance was evaluated on real-world Android malware from the TaintBench Suite.
    
    \item{\textbf{Knowledge-Enhanced Pre-trained Language Models KE-PLMs:}} These models represent a significant advancement in NLP by integrating structured external knowledge into the pre-training phase of language models ~\cite{yang2024survey,hu2023survey}. We can take advantage of these models for enhancing malware detection on source code.\\
    
    KE-PLMs incorporate external knowledge of various forms, including linguistic information, factual data, and domain-specific insights, to enrich the model's understanding and contextual awareness~\cite{zhen2022surveyKEPLM}. This approach enhances the model's ability to comprehend complex language structures and perform tasks requiring deep reasoning. KE-PLMs can perform better across a broad spectrum of NLP applications, contributing to more accurate and contextually relevant language processing outcomes by directly injecting external knowledge ~\cite{hu2023survey}.
 We believe that by incorporating structured external knowledge—such as information from knowledge graphs and domain-specific data into source code analysis, it is possible to identify relationships and behaviors associated with malware activities more effectively. Knowledge-Enhanced Pre-trained Language Models (KE-PLMs) can better capture these intricate patterns, leading to improved detection and understanding of malicious code.
    
    \item{\textbf{Large Concept Model (LCM) for malware code analysis}}
    Imagine a cybersecurity firm receives a suspicious file with potential malware. Traditional models typically analyze individual code snippets, often overlooking sophisticated obfuscation techniques. These limitations stem from token-level processing. In \cite{barrault2024large}, the authors introduced Large Concept Models (LCMs), which focus on concepts—complete semantic units—rather than tokens. This method allows for improved long-context understanding, abstract reasoning, and more efficient computation, ultimately enhancing performance in tasks such as cross-lingual and multimodal applications.
\end{itemize}

\subsection{Limitations and Challenges}
\begin{itemize}
\item{\textbf{Size of tokens for malware code decompiling:}}
Malware code analysis requires breaking down complex behaviors into smaller, manageable components for more precise and efficient examination. Due to the limitations of LLMs in processing lengthy prompts, generating code for individual unit functions—rather than for an entire malware program—enhances accuracy and clarity \cite{liu2024empirical}. This modular approach allows for better detection of malicious patterns, facilitates debugging, and improves adaptability to evolving threats. However, challenges remain in seamlessly integrating these unit functions and accurately simulating advanced malware techniques, such as obfuscation and evasion. Addressing these issues requires ongoing research and refinement in automated malware analysis. For example, in \cite{tan2024llm4decompile}, the LLM4Decompile-End model is limited by a maximum token length of 4,096 for decompiling.
 
\item{\textbf{Lack of familiarity with new malware patterns: }}
LLMs heavily rely on their training data, which restricts their effectiveness against novel or specialized malware patterns. If a model has not been fine-tuned or trained in specific new techniques, such as emerging malware variants, novel obfuscation methods, or unique software patterns, it may struggle to recognize or process these unfamiliar inputs. This limitation arises from the fact that LLMs rely on pre-existing patterns and examples within their training data. Consequently, when confronted with content or techniques outside of this training, the model may fail to accurately analyze or detect the relevant features, leading to potential gaps in analysis or missed patterns.\\\\
\end{itemize}

\section{Conclusion}

In this paper, we provided a comprehensive overview of the potential of large language models (LLMs) in the field of malware code analysis. Our focus was primarily on transformer-based models and their capabilities in detecting, classifying, and understanding malicious code. These advanced natural language processing (NLP) models have shown great promise in addressing modern cybersecurity challenges, particularly in identifying and mitigating evolving malware threats. \\ 

We discussed how LLMs can effectively analyze malicious code, leveraging their ability to understand patterns, detect anomalies, and classify threats with high accuracy. By harnessing the power of deep learning and pre-trained models, LLMs have the potential to revolutionize the field of malware analysis, making it more efficient and scalable. Additionally, we explored various publicly available datasets that are instrumental in malware research, especially for tasks such as pre-training and fine-tuning LLMs. These datasets play a crucial role in static malware analysis, enabling researchers and cybersecurity professionals to train models on real-world malicious code samples.\\  

Furthermore, we examined how the integration of LLMs into malware detection and mitigation strategies can lead to more proactive cybersecurity measures. By automating aspects of malware classification and code analysis, LLMs enhance detection speed, reduce human effort, and improve overall security posture. Their ability to adapt to new and evolving malware threats makes them a valuable tool in cybersecurity defence mechanisms.\\ 

Despite the current advancements, we believe that LLMs hold significant potential for further improving malware analysis methodologies and strengthening cybersecurity defences. As these models continue to evolve, they could play a pivotal role in shaping the future of threat intelligence, malware detection, and cybersecurity automation.

\printcredits
\bibliographystyle{IEEEtran}
\bibliography{references}

\end{document}